\documentclass[preprint,12pt]{elsarticle}

\usepackage{epsfig}

\usepackage{amssymb}
\usepackage{amsthm,amsmath,bbm}

\usepackage{psfrag}

\newcommand{\bl}{}

\usepackage[usenames,dvipsnames]{xcolor}

\journal{Mathematical Biosciences}

\begin{document}


\begin{frontmatter}



\title{A unified model for wild resistance dynamics and weed control using herbicide}


\author[usp]{L. H. B. Bertolucci}
\ead{bertolucci@icmc.usp.br}

\author[usp]{E. F. Costa}
\ead{efcosta@icmc.usp.br}

\author[usp]{V. A. Oliveira}
\ead{vilma@sc.usp.br}

\author[emb]{D. Karam}
\ead{decio.karam@embrapa.br}

\address[usp]{Universidade de Sao Paulo, Sao Carlos, Brazil}

\address[emb]{Empresa Brasileira de Pesquisa Agropecuaria,
Embrapa Milho e Sorgo, Sete Lagoas, Brazil}

\begin{abstract}
A major issue in the modelling of weed resistance to herbicide lies in effectively handling
the wild dynamics, that is, the allele frequency prior to the herbicide application, 
and in particular when starting to use the herbicide.
The wild allele frequency is a key variable as the resistance evolution is highly sensitive to it, moreover it 
is extremely difficult to measure in the agricultural field. 
In this paper we propose a model for weed control  
that handles the allele frequency in a direct manner, 
grounded on the very dynamics of the weed life cycle, 
with no need of a priori distributions nor the use of the Hardy-Weinberg equilibrium. 
The proposed model is individual based, stochastic, and considers some 
phenomena like the 
relative fitnesses and mutation that are prominent in the resistance dynamics without herbicide.
A case study is presented for the herbicide nicosulfuron in a field with  the weed {\it Bidens pilosa}.
Another two models having a standard deterministic dynamics are compared with ours in terms 
of the initial allele frequency, its time evolution, and the resistance visualization in the field, 
indicating that the proposed model is effective to provide more realistic simulations for the weed resistance.
\end{abstract}

\begin{keyword} stochastic model \sep individual-based model \sep herbicide resistance \sep
allele frequency dynamics \sep weed control.
\end{keyword}

\end{frontmatter}


\section{Introduction}
\label{lab:intro}

Weed resistance to herbicide is a key issue in food production
as it promotes efficiency loss of herbicide, one of the most worldwide
used strategy to control weeds  \cite{Jasieniuk_1996}.  
Due to the difficulty and high cost
in the development of new pesticides, much effort has been made in
understanding the multiple factors surrounding this phenomenon.
In this context, it is well recognized that computer simulation modelling can
help understanding the relationship between the use of herbicide and
its resistance evolution \cite{Renton_etal2014}.
During the last decades many models have been proposed 
to describe and predict the herbicide resistance evolution 
\cite{Renton_etal2014,Holst_etal_2007}.

\par

The available dynamic models are, to our knowledge, concerned with
the resistance evolution in short term, typically in a time interval 
$[0\;\; T]$, being the order of magnitude of $T$ less than 2.  
A notorious difficulty that arises when dealing with these 
models is to accurately estimate the resistant allele frequency at $t=0$, 
and in a more general context, the allele frequency dynamics for pre-herbicide 
application  periods, which we refer to as the \emph{wild resistant allele frequency dynamics}, 
or \emph{wild frequency} for short. 
According to \cite{Bagavathiannan_etal_2013}, the wild frequency
is critical for studying the resistance risk. 
Similar discussions were also presented in
\cite{Neve_etal_2011,Thornby_etal_2009}.
Most of simulation results assume this quantity as deterministic, see e.g. 
\cite{Thornby_etal_2009,Cavan_2000,Diggle_etal_2003,Neve_etal_2003_I,Neve2008,Renton2011,Manalil2012}. 
However, as the natural evolution of resistance is mainly driven
by both mutation and natural selection (specifically, 
the relative fitnesses between resistant and susceptible individuals),
two phenomena that are largely known as random ones, 
it is important to consider the wild frequency as a random variable,
as in 
\cite{Bagavathiannan_etal_2013,Neve_etal_2011,Neve_etal_2003_II,Neve_etal_2011B}. 
The main difficulty in this case is to estimate its probability distribution, 
once it is extremely hard to measure this variable in agricultural 
fields to collect the necessary data 
\cite{Renton_etal2014,Neve_etal_2011,Neve_etal_2003_I,Neve2008}.


\par

In this paper we propose a stochastic and individual-based model 
for the resistance evolution that handles the wild frequency 
issue in a simple, unified manner. 
In fact, we use a single dynamic model in a large time 
interval $[-T_0\;\; T]$ with 
$T_0,T>0$, and $T_0$ large enough such that the initial condition at time $t=-T_0$  
has almost no relation with the seed bank state at $t=0$, 
when the application of herbicide starts. 
The main benefit in using this strategy is that there is no need 
to define a priori distribution for the wild resistant allele frequency
and also for the use of the Hardy-Weinberg equilibrium 
\cite[Chapter 3]{Smith_Book_1998}.
Using the model in the interval $[-T_0\;\; 0]$
it is also possible to estimate the distribution of wild frequency, 
called {\it wild distribution},
as performed in Section \ref{sec:Res:semherb}, see 
Figure \ref{fig:FreqAllele2}. 
Moreover, the model allows us to study the allele frequency distributions for the 
post-herbicide application period $[0\;\; T]$, 
also performed in Section \ref{sec:Res:compModels}.

\par 

The proposed model is based on the life cycle of 
a weed population considering a post-emergent herbicide \cite{Radosevich_etal_BOOK_2007,Zimdahl_BOOK_2013}, involving 
the germination in soil, emergence, herbicide application, 
flowering, seed production, 
mortality of seeds in soil and again germination.
We employ a dose response function obtained from greenhouse experiments,
relating the dose response of susceptible and 
resistant \textit{Bidens pilosa} to the application of the nicosulfuron herbicide. 
The models also allow to emulate competition with other species and other 
natural factors that 
prevent a booming population, in a simple way, which can be interpreted 
as if an ``alternative herbicide'' were employed for $t<0$. 
The stochastic model employs binomial distributions \cite{degroot2002probability}
in the most random processes involved in the life cycle of weeds,
including the mutation phenomena 
and relative fitnesses. 
It is worth noting that the use of binomial distributions is suggested 
by \cite{Renton_etal2014} and is also employed in \cite{Renton2011}.
A contribution of this paper is to combine all the above ingredients in a single model, 
which involved integration of existing equations and development of some new ones, 
such as the stochastic reproduction equation in \eqref{eq:totalProd}, and to present the complete 
equations in a detailed manner.

\par

Another contribution is to compare the proposed 
model with a deterministic model and also with a simple extension of it, which we 
refer to as the \emph{hybrid model}, defined 
by a deterministic dynamics for $t\geq 0$ and a random initial condition at $t=0$.
The motivation behind the deterministic model lies in the fact that it 
is able to estimate averages of the main variables, has low computational burden
and is very popular \cite{Renton_etal2014}, so it is important to assess the quality of its estimates.   
As for the hybrid model, we check how the deterministic 
dynamics act on the random initial condition, by 
simulating the distributions and comparing with the 
ones given by the stochastic model. 
The simulation results presented here focus on the resistance risk in terms 
of visualization of resistance in the field, and also on the 
evolution of the wild distribution when the 
application of herbicide starts.

\par 
The text is organized as follows.
In Section \ref{lab:model} we present the stochastic, hybrid and 
deterministic models.
In Section \ref{sec:result} we present  estimates for  
the wild frequency and for the allele distributions
in the post herbicide application period. 
We also present a comparison between the models in terms
of resistance visualization index.
At the end of this section we carry out a discussion about the results. 
Finally, some conclusion are made in Section \ref{sec:conclusions}.

\subsection*{Notation}

Throughout the paper capital letters indicate a vector or 
some of its elements and lowercase letters  indicate scalar variables
or parameters. 
We reserve boldface letters for stochastic variables and standard letters 
to deterministic ones.
The symbols $\mathbb N$ and $\mathbb R$ represent the set of natural 
and real numbers, respectively.
We use $\mathcal E\{.\}$ for the expected value of a stochastic variable,
$\mathcal B(n,p)$ for a binomial distribution \cite{degroot2002probability} with 
$n$ being the number of experiments 
and $p$ its probability of success;
$\mathcal B^M(n,P)$ is employed for the multinomial distribution \cite{degroot2002probability}, 
with $n$ already been defined 
and $P$ the probability vector, and $\mathcal N(m,s^2)$
for the normal distribution \cite{degroot2002probability}
with $m$ being the mean and $s$ the standard deviation.

\section{Weed population models}
\label{lab:model}

The considered models are based on the life cycle of a weed population
and post-emergent herbicides.
We assume that the population is closed, that is, there is no gene exchange 
from surrounding agricultural fields.
The Figure 1 presents all variables employed by the models and indicates
how they are related to each other.  Note that the time sequence is
indexed to the weed life cycle. We focus on the resistance
evolution of a certain herbicide referred to as the
\emph{target herbicide}, denoted $u_t$. It is worth noting that a 
second herbicide is included in the models, denoted $v_t$,
providing more flexibility, for instance, allowing to
simulate the use of other herbicides or even emulate natural
population control; we refer to this herbicide as the 
\emph{alternate herbicide}, and assume it causes no selective pressure regarding
resistance to the target one.

\par

\begin{figure}
\label{fig:LyfeCycle}
\centering
\def\svgwidth{12cm}
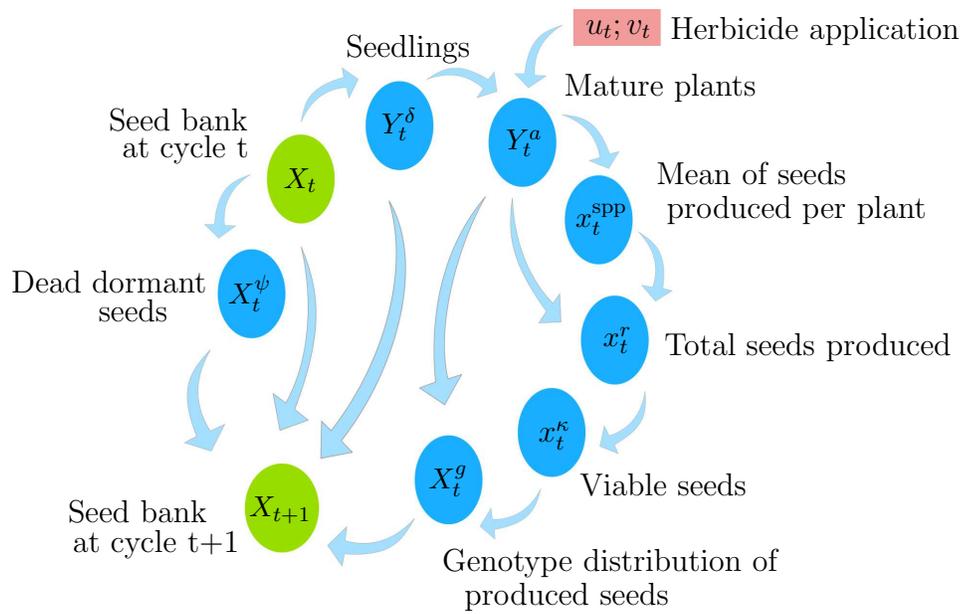
\caption{Flow chart of weed life cycle considered by the models. 
The disks indicate the employed variables.
The value or distribution of the variables at the head of an arrow is a function of 
the variables at the tail.
The green disks highlight the state variables in two consecutive cycles, $t$ and $t+1$.}
\end{figure}
\par

\subsection{Stochastic model}

Following papers in weed resistance such as  
\cite{Diggle_etal_2003,Neve_etal_2003_I,Neve_etal_2003_II}, 
we suppose that the weed resistance to the target herbicide is associated  
to a gene arranged at a 
single genetic locus. We also suppose that the weed resistance is due to 
dominant alleles which are denoted by the capital letter $A$. Therefore, the  
list of all possible genotypes in the weed  population is given by 
$\mathcal{G} = \begin{pmatrix} AA & Aa & aa \end{pmatrix}$.
For each genotype, we denote the quantity of seeds in the seed bank
at instant $t$ as 
${\bf X}_{t}=\begin{bmatrix}{\bf X}_{1,t}\; & {\bf X}_{2,t}\; & {\bf X}_{3,t} 
\end{bmatrix}'$,
where  ${\bf X}_{t}\in \mathbb N^3$ and ${\bf X}_{i,t}$ the amount the $i$-th
genotype seed in list $\mathcal{G}$. As an example,  ${\bf X}_{2,5}$ is the amount 
of seed type $Aa$ at instant $5$.

\par

Assuming that each seed in the area has an uniform probability in time
to germinate, referred here as constant probability, the amount of
seedling is a random variable with the following probability,
 \begin{equation}
{\bf Y}^\delta_{i,t}\sim \mathcal B({\bf X}_{i,t},\delta)
\end{equation} 
where ${\bf Y}^\delta_{i,t}\in\mathbb N$ is the amount of seed of
genotype $i$ germinated in the $t-th$ cycle and 
$\delta$ is the germination  probability of
each seed.  We mention that parameter uncertainty and time variation
can be incorporated in the model, e.g. by considering 
$\delta$ as a time varying parameter in the above probability distribution.


Assuming that the number of mature weeds remaining from the
previous cycle is negligible\footnote{This is valid for instance when
the production cycle is sufficiently short so that there is only one
generation of mature weed per cycle.  If this hypothesis is void, 
the remaining weeds can be added to ${\bf Y}^{a}_{i,t}$
in \eqref{eq:adultas}.}, we have that the total of
mature weeds observed at time instant $t$, denoted by 
${\bf Y}^{a}_{t}\in\mathbb N^3$, obeys the following distribution,
\begin{equation}\label{eq:adultas}
{\bf Y}^{a}_{i,t} \sim \mathcal B\left({\bf Y}^\delta_{i,t}, \xi_i
M_{i}(u_t,v_t)\right),
\end{equation}
where $\xi_i M_{i}(u_t,v_t)$ is the probability that a germinated 
weed survives until the mature stage, addressed as follows.  
We assume that the herbicides are not
employed together in a same cycle, so that the survival probability to the herbicide can
be modelled by the functions
$M_i(u,v)= (1-\rho^{R}(u))(1 - \rho(v))$, for $i=1,2$ and 
$M_3(u,v)= (1-\rho^{S}(u))(1 - \rho(v))$,
%
where $\rho^{R}(u)$ and $\rho^{S}(u)$ are the expected mortality of
resistant and susceptible seedlings, respectively, due to the target
herbicide application, and $\rho(v)$ is the expected mortality due to
the alternative herbicide. Note that the uses of $v$ 
doesn't cause selective pressure.
Now, to account for the natural mortality 
we define $\xi_i = \gamma(1-c)$, for $i = 1,2$
and $\xi_3 = \gamma $, where $\gamma$ is the natural survival of weeds
and $c$ is the fitness cost parameter,  
considered in several papers as in \cite{RouxReboud2007,Roux_etal2008,Neve_2008}.
We have included the fitnees cost because there is usually a metabolic penalty
associated with the resistance to a certain herbicide \cite{Roux_etal2008}, 
leading to decrease in the reproductive success.
%

\par 


The total generated seeds
can be obtained by ${\bf y}^a_{t} {\bf  x}^{\text{spp}}_{t}$, 
where ${\bf y}^a_{t}$ is the number of total adult plants, that is,
${\bf y}^a_{t} = \sum_{i=1}^3 {\bf Y}^{a}_{i,t}$, and 
${\bf x}^{\text{spp}}_{t}\in \mathbb R$ is the mean 
of seeds produced per plant at $t$.
Since we do not have the probability
distribution for ${\bf x}^{\text{spp}}_{t}\in \mathbb R$ (and have
found no literature on this subject), we adopt a normal distribution
\begin{equation}
\label{eq:totalProd}
{\bf y}^a_{t} {\bf x}^{\text{spp}}_{t} \sim N\left({\bf
  y}^a_{t}x^{\text{spp}}_{t}, \left(\varphi {\bf y}^a_{t}
{x}^{\text{spp}}_{t} \right)^2 \right),
\end{equation}
where ${x}^{\text{spp}}_{t}:= \mathcal E\{{\bf x}^{\text{spp}}_{t}|{\bf y}^a_{t}\}$
and $\varphi$ is a parameter to adjust the standard deviation of the distribution.
To calculate ${x}^{\text{spp}}_{t}$ we use the following expression,
\begin{equation}
\label{eq:meanSeeds}
{x}^{\text{spp}}_{t} =
\left\{
  \begin{array}{lr}
    \frac{F g} {F + g\,{\bf y}^a_{t}/G} & : {\bf y}^a_{t} > 0\\ 
        0 & :  {\bf y}^a_{t} = 0.
  \end{array}
\right\},
\end{equation}
where $F$ is the agriculture field area in square meters, and $g$ and $G$
are bounds related to ${x}^{\text{spp}}_{t}$ in the sense that 
${x}^{\text{spp}}_{t}$  converges to $g$ when  ${\bf y}^a_{t}\rightarrow 0$ and 
$\frac{{\bf y}^a_{t}{x}^{\text{spp}}_{t}}{F}$ converges to $G$ 
when ${\bf y}^a_{t}\rightarrow \infty$. 
Equation \eqref{eq:meanSeeds} was adapted from \cite{Neve_etal_2011,Neve2008}, 
and takes into account the
intra-specific competition in such a manner that ${x}^{\text{spp}}$
decreases with the increase of the adult plants. 
It is worth to highlight that in the cited papers the equation is slightly 
different, its output is the amount of seeds produced per square meter.
Now, rounding down the values generated by the normal distribution,
we obtain the total produced seeds as  ${\bf x}^r_{t} =
\lfloor {\bf y}^a_{t} {\bf x}^{\text{spp}}_{t} \rfloor$.  
Also, considering that each seed has a probability $\kappa$ of becoming a viable
seed, we get
\begin{equation}
{\bf x}^{\kappa}_{t} \sim \mathcal B\left({\bf x}^r_{t},\kappa \right),
\end{equation}
where ${\bf x}^{\kappa}_{t} \in \mathbb N$ is the total viable seeds produced. 
Finally, we determine the number of viable seeds of each genotype, 
denoted by ${\bf  X}^{g}_{t} \in \mathbb N^3$, employing the following distribution,
\begin{equation}
\label{eq:genseed}
{\bf X}^{g}_{t} \sim \mathcal B^M \left ({\bf x}^{\kappa}_{t}, P(G_t)\right),
\end{equation}
where $P(G_t) = [P(G_{1,t})~P(G_{2,t})~P(G_{3,t})]'\in\mathbb{R}^3$,
with $P(G_{i,t})$ being the probability of a generated seed to have
genotype $i$. The calculation of $P(G_{i,t})$, presented in \ref{sec:app:PG},
depends on the quantity of mature plant of each genotype ${\bf Y}^{a}_{i,t}$, 
on the probability of mutation in the gametogenesis, denoted  $P(m)$,
and on the probabilities of a seed be generated by self or of cross-fertilization, 
denoted $P(F_a)$ and  $P(\bar F_a)$, respectively.

To model the mortality of dormant seeds we use the following distribution,
\begin{equation}
\label{eq:germ}
  \begin{array}{lr}
{\bf X}^{\psi}_{i,t} \sim \mathcal B({\bf X}_{i,t}-{\bf Y}^\delta_{i,t},\psi).
  \end{array}
\end{equation}
where $\psi$ is the seed death probability.
Finally, using balance of seeds, we obtain the characterization 
of the seed bank for all $t\geq t_0$ using the following recursive equation,
\begin{eqnarray}
\Bigg\{
 \begin{array}{ccl}
{\bf X}_{i,t+1} &=& {\bf X}_{i,t} -{\bf Y}^\delta_{i,t} - {\bf
  X}^{\psi}_{i,t} + {\bf X}^{g}_{i,t}\;, \\
{\bf X}_{i,t_0} &=& \breve{\bf X}_{i}\;,
  \end{array}
\end{eqnarray}
where $t_0$ is a predetermined starting time and $\breve{{\bf X}}_{i}$
is a stochastic initial condition for the seed bank.

\subsection{Deterministic model}

In this section, we describe the deterministic model.
This model takes into account the same stages of weed life cycle 
considered for the stochastic model. For conciseness, we present
the equations directly, as following
\begin{eqnarray}
\label{eq:ModDetEQ1}
 Y^\delta_{i,t} &=& \delta  X_{i,t}, \\
 Y^{a}_{i,t} &=&  \xi_i M_{i}(u_t,v_t)  Y^\delta_{i,t}, \\
x^{\text{spp}}_{t} &=& \Bigg\{
  \begin{array}{lr}
    \frac{F g} {F + g\,y^a_{t}/G} & : y^a_{t} >  0, \\
    0 & :  y^a_{t} = 0,
  \end{array}\\
 x^{\text{r}}_{t} &=&    y^a_{t}  x^{\text{spp}}_{t}, \\
 x^{\kappa}_{t} &=& \kappa  x^{\text{r}}_{t}, \\
 X^{g}_{t} & =&   x^{\kappa}_{t} P(G_t),\\
\label{eq:ModDetEQend}
 X^{\psi}_{i,t} &=& \psi\left( X_{i,t}- Y^\delta_{i,t}\right).
\end{eqnarray}
%
%
Using the seed balance, we obtain 
\begin{eqnarray}
\Bigg\{
 \begin{array}{ccl}
 X_{i,t+1} & = &\left(1 -\delta -\psi +\psi\delta\right)  X_{i,t} 
+  X^{g}_{i,t}\quad , \\
 X_{i,t_0} & = & \breve{X_{i}},
 \end{array}
\end{eqnarray}
where $\breve{X}_{i}$ is a deterministic initial condition for the seed bank.

The stochastic and deterministic models are similar only in terms of 
number of variables and
equations, as they are derived from the same life cycle stages of a
weed population; the parameter $\varphi$, in Equation \eqref{eq:totalProd}, 
is the only additional one in the stochastic
model.  Apart from this aspect, the models are quite different.  One
of the major difference lies in the domains of variables.  The state
variables of the stochastic model take values in the set of integer
numbers, which seems more realistic than the real variables of the
deterministic model.  Also, the interpretation of some parameters is
different; e.g. $\delta$ is the probability that a seed in the bank
germinate at a cycle in the stochastic model, while it is the fraction
of seeds that germinate in the deterministic one, leading to a exact
(real) number of $\delta { \bf X}_{i,t}$ germinated seeds, which seems
less consistent with the real-world application.  Of particular
relevance is the case, when the number of seeds is small, as frequently
found when handling small areas, the number of seeds of resistant
genotype turns out to be smaller than in the deterministic model.
These fractions of seeds propagate in this model as if they were able
to germinate and produce new plants.

It is worth to note that the deterministic equations cannot be obtained 
by taking expectation of the stochastic ones. 
In fact, as the expectation does not commute with some operators, we have that the expectation 
does not ``move freely'' between equations and inside them.  
To illustrate, while in the deterministic model 
we assumed that ${ x}^r_{t} =  { y}^a_{t}  { x}^{\text{SPP}}_{t}$,
from \eqref{eq:totalProd}, we find
\begin{equation}
 \mathcal E\{{\bf y}^a_{t}  {\bf x}^{\text{SPP}}_{t}\}
\not= \mathcal E\{{\bf y}^a_{t}\} \mathcal E\{{\bf x}^{\text{SPP}}_{t}\},
\end{equation}
because the number of mature weeds and mean 
of seeds produced per plant are correlated.

%
%


\subsection{Hybrid model}
Simplicity is one of the main features of the deterministic model,  
perhaps explaining its popularity in the weed resistance literature. 
On the other hand, the simplicity affects the comparisons with the stochastic model; for instance, 
the distribution of the allele frequency is trivial 
(concentrated in a point) making superficial the comparison with the 
relatively complex distribution given by the stochastic model 
(see e.g. Figure \ref{fig:FreqAllele2}).
With this motivation, we introduce a third model, which we refer to 
as the hybrid model, which preserves the deterministic dynamic, \eqref{eq:ModDetEQ1}
- \eqref{eq:ModDetEQend}, 
but with a random initial condition
\begin{eqnarray}
\Bigg\{
 \begin{array}{ccll}
 X_{i,t+1} & = &\left(1 -\delta -\psi +\psi\delta\right)  X_{i,t} 
+  X^{g}_{i,t}, &\quad \textit{for} \quad t\geq t_s\; \\
 X_{i,t_s} & = & \breve{{\bf X}}_{i},
 \end{array}
\end{eqnarray}
where $t_s$ is the starting time and 
 $\breve{\bf X}_{i}$ is a stochastic initial condition.


\section{Simulation results}
\label{sec:result}
We start with the simulation results for the pre target herbicide 
application period, $[-T_0 \; 0]$, 
with emphasis on the wild frequency 
and its distribution according to the stochastic model.
Then we turn our attention to the herbicide application period $[0 \; T]$. 
Recall that the application of target herbicide is zero ($u_t=0$)  
in the interval $-T_0\leq t \leq -1$, and the alternative herbicide is
applied at fixed doses $v_t=60 \text{\,g\,ha}^{-1}$ 
(please, see \ref{sec-param} for motivation) 
to emulate the usage of
any other herbicide, competition with other species and other
factors preventing a booming weed infestation prior to the first usage of the 
target herbicide. 
Conversely, we set $v_t=0$ and $u_t= 60 \text{\,g\,ha}^{-1}$ for $0\leq t \leq T$.

\par

The parameters adopted in all simulations are given in \ref{sec-param}.
We denote the seed trajectory generated by the 
deterministic model by $\{ X_t,-T_0\leq t \leq T\}$ 
and, for the $k$-th realization of a Monte
Carlo simulation 
we denote the seed trajectories of the stochastic and hybrid models by 
$\{{\bf  X}_t^{\omega_k},-T_0\leq t \leq T\}$ and
  $\{\mathfrak{X}_t^{\omega_k}, 0\leq t \leq T \}$, respectively.
The Monte Carlo simulation was performed with $10^4$ realizations.
The initial seed bank (at $t=-T_0$) for the stochastic model is set to $2200$
seeds $m^{-2}$, with no resistant seeds.
We recall that we set the initial allele frequency of hybrid model using the stochastic
model outputs,
$\{\mathfrak{X}_0^{\omega_k} = {\bf  X}_0^{\omega_k}, 1\leq k \leq 10^4 \}$. 
It made possible to compare the stochastic and hybrid model output
on the evolution and visualization of resistance
for the post herbicide application period.


\subsection*{\bf Pre target herbicide  application period and wild distribution}
\label{sec:Res:semherb}

Trajectories of the resistant allele frequency and the seed bank have been 
computed based on ${\bf  X}_t^{\omega_k}$ and $X_t$ using the equations given in 
\ref{sec:app:AE}, and they are illustrated in Figure 1. 
We plot only the estimated average and 
standard deviation linked with the stochastic model as plotting
$10^4$ realizations is useless. 
The allele frequency estimated by the Mutation-Selection Theory,
given by $P(m)/c$ \cite[Chapter 4]{Smith_Book_1998}, 
is also included in Figure 1. 
Note that $\mu_{{ X}_t}$ and $\overline{\mu}_{{\bf X}_t}$ at $t=0$ 
are close to $P(m)/c$, as expected.
These results strongly suggest that the stochastic model is stable, 
as unstable dynamics usually leads to divergence of the average or the 
standard deviation.

\par

\begin{center}
\begin{figure}[!ht]
\begin{tabular*}{0.75\textwidth}{ c c }
          \psfrag{10}[c]{\footnotesize 10}
          \psfrag{-10}[r][r]{\scriptsize -10}
          \psfrag{-9}[r][r]{\scriptsize -9}
          \psfrag{-8}[r][r]{\scriptsize -8}
          \psfrag{-7}[r][r]{\scriptsize -7}
          \psfrag{-6}[r][r]{\scriptsize -6}
          \psfrag{-5}[r][r]{\scriptsize -5}
          \psfrag{-4}[r][r]{\scriptsize -4}
          \psfrag{-1000}[t][c]{\footnotesize -1000}
          \psfrag{-800}[t][c]{\footnotesize -800}
          \psfrag{-600}[t][c]{\footnotesize -600}
          \psfrag{-400}[t][c]{\footnotesize -400}
          \psfrag{-200}[t][c]{\footnotesize -200}
          \psfrag{0}[t][c]{\footnotesize 0}
          \psfrag{L E}{\small $\overline{ \mu}_{{\bf X}_t}$}
          \psfrag{L ES}{\small $\widehat{\mu}_{_{{\bf X}_t}} $}
          \psfrag{L D}{\small $ \mu_{X_t}$}
          \psfrag{L MXXX}{\small $P(m)/c$}
          \psfrag{Allele Frequency}[b][c]{Allele frequency}
          \psfrag{t}[t]{t}
          \includegraphics[height=6.0cm]{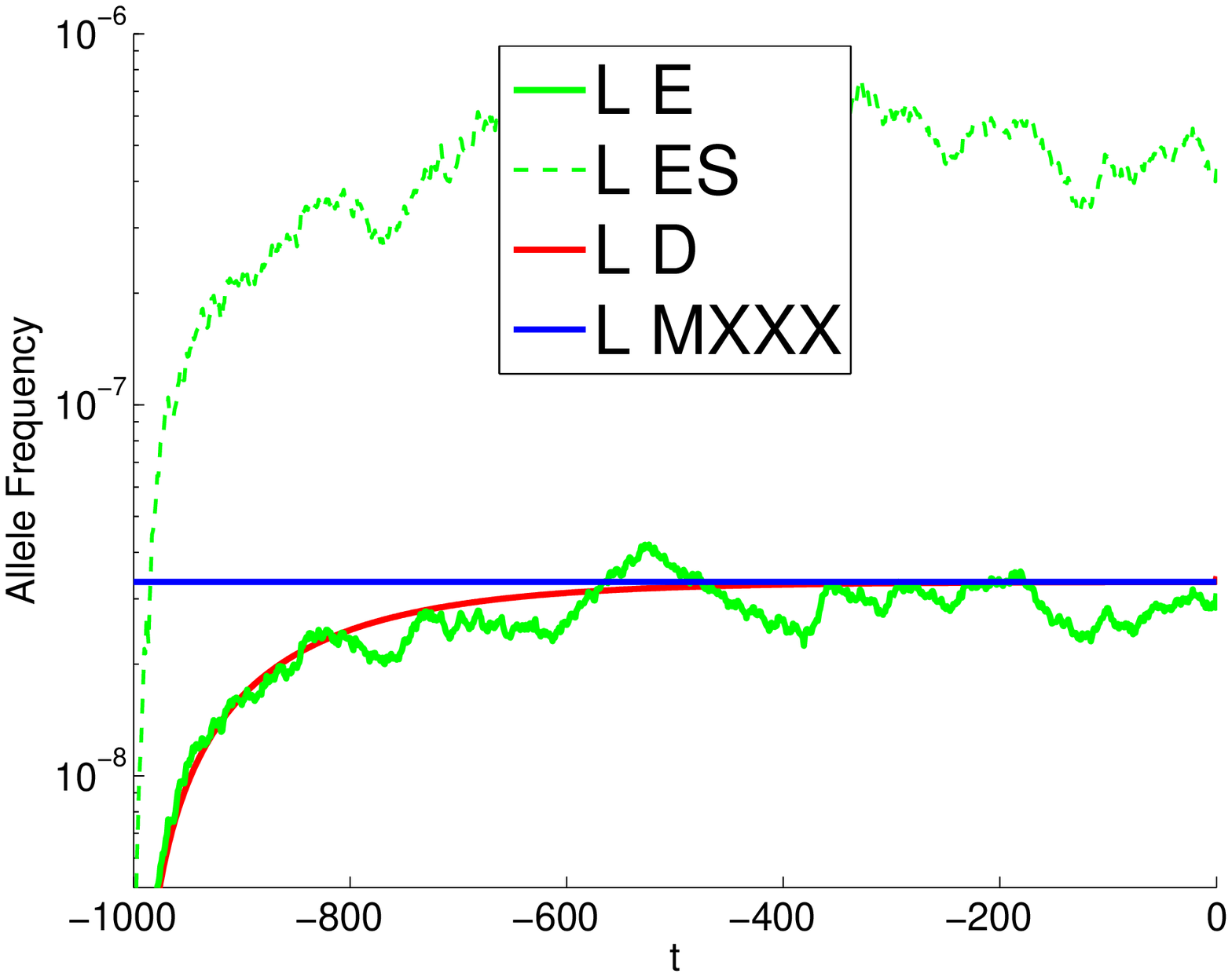} &
          \psfrag{2400}[r][r]{\scriptsize 2400}
          \psfrag{2300}{}
          \psfrag{2200}[r][r]{\scriptsize 2200}
          \psfrag{2100}{}
          \psfrag{2000}[r][r]{\scriptsize 2000}
          \psfrag{1900}{}
          \psfrag{1800}[r][r]{\scriptsize 1800}
          \psfrag{1700}{}
          \psfrag{1600}[r][r]{\scriptsize 1600}
          \psfrag{1500}{}
          \psfrag{1400}[r][r]{\scriptsize 1400}
          \psfrag{-1000}[t][c]{\footnotesize -1000}
          \psfrag{-900}[t][c]{}
          \psfrag{-800}[t][c]{\footnotesize -800}
          \psfrag{-700}[t][c]{}
          \psfrag{-600}[t][c]{\footnotesize -600}
          \psfrag{Seeds.m2}[b][c]{Seeds $m^{-2}$}
          \psfrag{L EX}{\small $\overline{\eta}_{{\bf X}_t}$}
          \psfrag{L ES}{\small $\widehat{\eta}_{{\bf X}_t}$}
          \psfrag{L DX}{\small $ \eta_{X_t}$}
          \psfrag{t}[t]{t}
          \includegraphics[height=6.0cm]{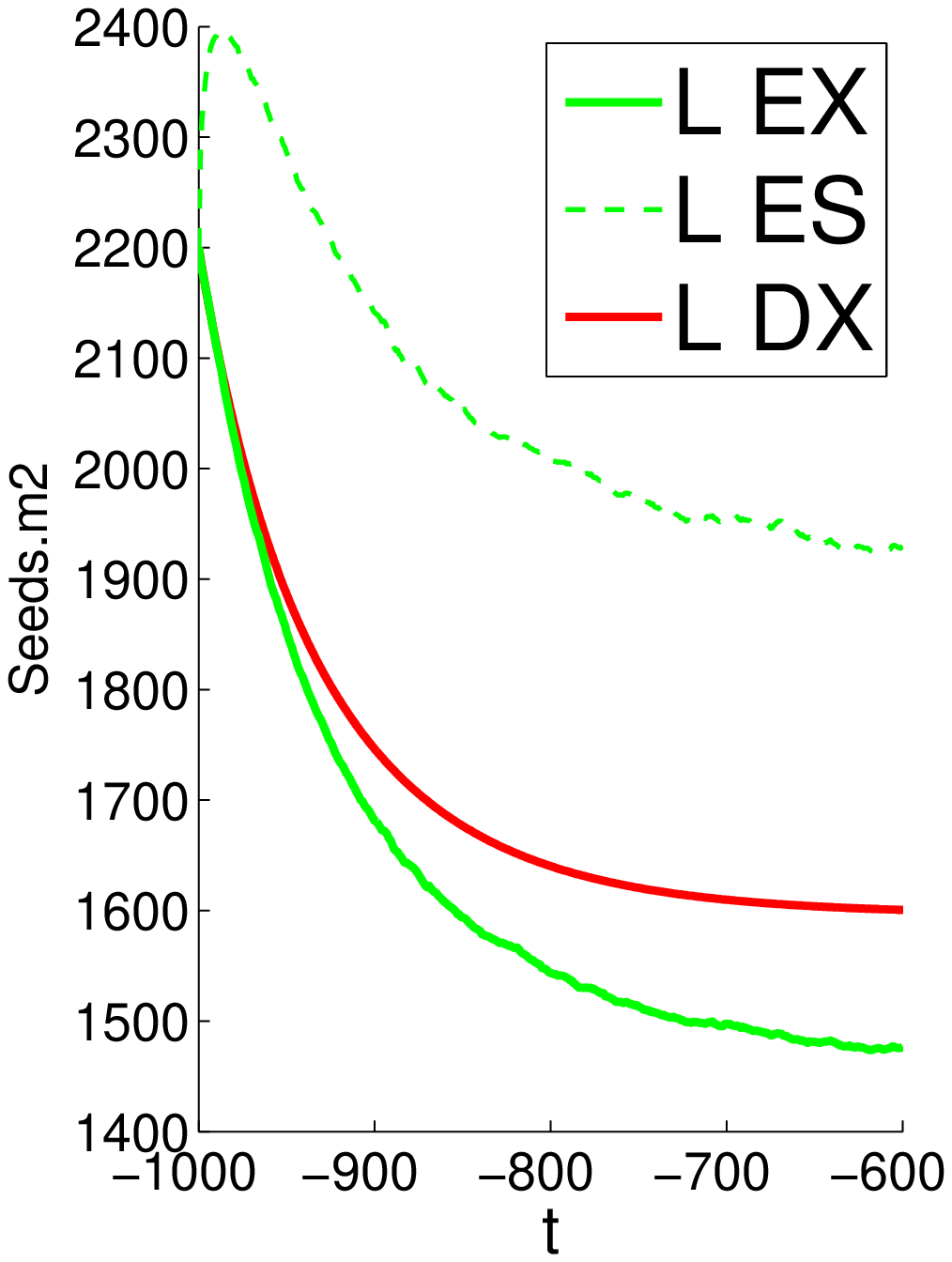} \\
\end{tabular*}
\caption{
Allele frequency and seed bank given by the deterministic model (red line), 
the average and standard deviation given by 
the stochastic model  (solid green and dashed green, respectively).
The allele frequency given by the Mutation-Selection Theory is 
expressed by blue line.}
\label{fig:FreqXSeedDensity}
\end{figure}
\end{center}

\par

\begin{figure}[!ht]
\begin{tabular*}{0.75\textwidth}{ c c }
\psfrag{Media}{\small $\mathbf \mu^{\rho}({ \mathcal X}_T)$}
\psfrag{Ocorrencias}[b][c]{\small Occurrences}
\psfrag{Frequencia do Alelo}[t][c]{\small Allele Frequency}
\psfrag{Ocur}{${n}^{\rho}({ \mathcal X}_T)$}
\psfrag{100}[t][t]{\footnotesize $0$}
\psfrag{10}[0.9]{\footnotesize $10$}
\psfrag{0}[tc][t][0.6]{\footnotesize $0$}
\psfrag{1}[tc][t][0.6]{\footnotesize $1$}
\psfrag{2}[tc][t][0.6]{\footnotesize $2$}
\psfrag{3}[tc][t][0.6]{\footnotesize $3$}
\psfrag{4}[tc][t][0.6]{\footnotesize $4$}
\psfrag{-10}[tc][t][0.6]{\footnotesize $-10$}
\psfrag{-9}[tc][t][0.6]{\footnotesize $-9$}
\psfrag{-8}[tc][t][0.6]{\footnotesize $-8$}
\psfrag{-7}[tc][t][0.6]{\footnotesize $-7$}
\psfrag{-6}[tc][t][0.6]{\footnotesize $-6$}
\psfrag{-5}[tc][t][0.6]{\footnotesize $-5$}
\psfrag{-4}[tc][t][0.6]{\footnotesize $-4$}
\includegraphics[scale=0.35]{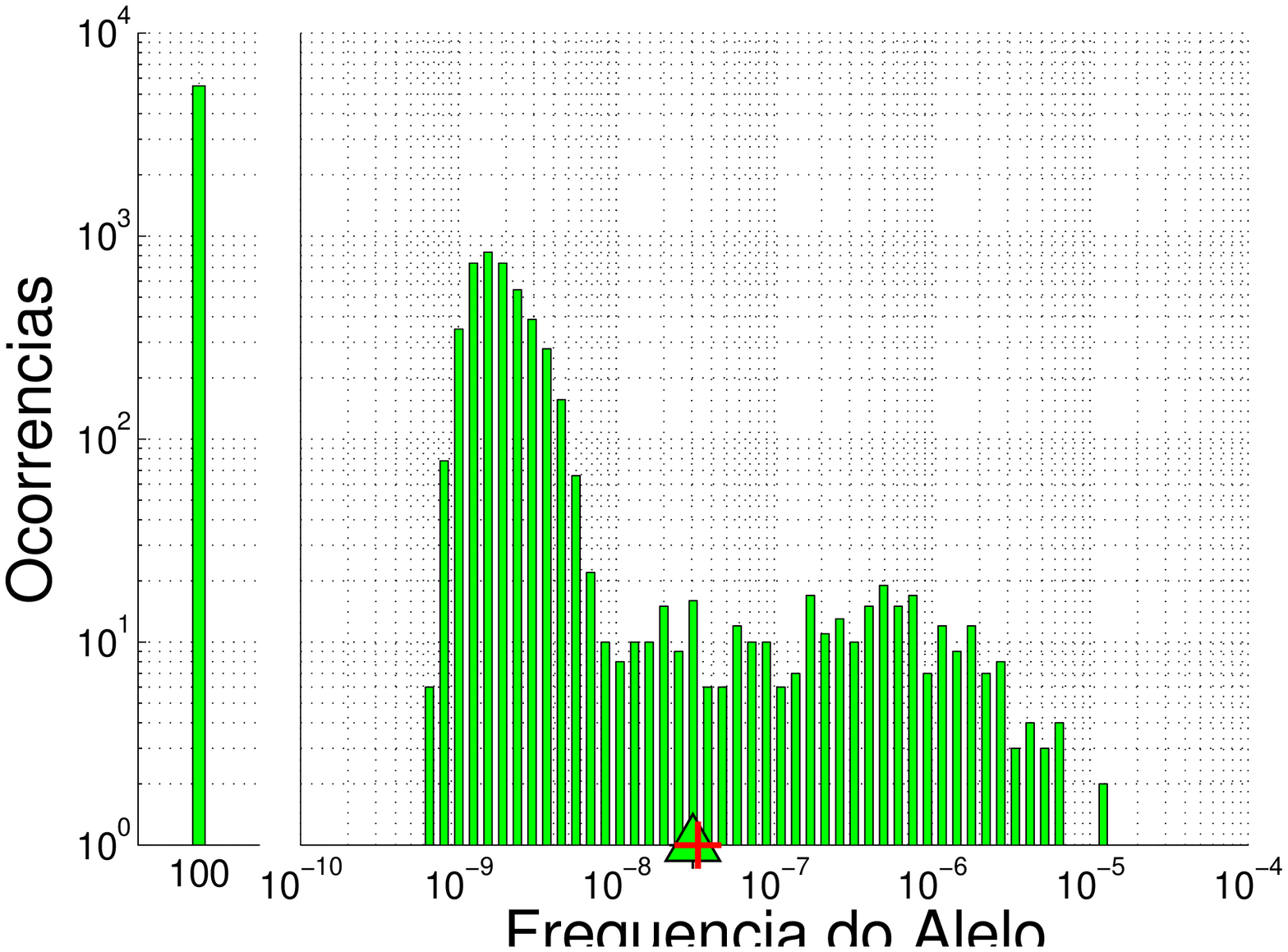} &
\psfrag{Media}{\small $\mathbf \mu^{\rho}({ \mathcal X}_T)$}
\psfrag{Ocorrencias}[b][c]{\small Occurrences}
\psfrag{Frequencia do Alelo}[t][c]{\small Allele Frequency}
\psfrag{Ocur}{${n}^{\rho}({ \mathcal X}_T)$}
\psfrag{100}[t][t]{\footnotesize $0$}
\psfrag{10}[0.9]{\footnotesize $10$}
\psfrag{0}[tc][t][0.6]{\footnotesize $0$}
\psfrag{1}[tc][t][0.6]{\footnotesize $1$}
\psfrag{2}[tc][t][0.6]{\footnotesize $2$}
\psfrag{3}[tc][t][0.6]{\footnotesize $3$}
\psfrag{4}[tc][t][0.6]{\footnotesize $4$}
\psfrag{-10}[tc][t][0.6]{\footnotesize $-10$}
\psfrag{-9}[tc][t][0.6]{\footnotesize $-9$}
\psfrag{-8}[tc][t][0.6]{\footnotesize $-8$}
\psfrag{-7}[tc][t][0.6]{\footnotesize $-7$}
\psfrag{-6}[tc][t][0.6]{\footnotesize $-6$}
\psfrag{-5}[tc][t][0.6]{\footnotesize $-5$}
\psfrag{-4}[tc][t][0.6]{\footnotesize $-4$}
\includegraphics[scale=0.35]{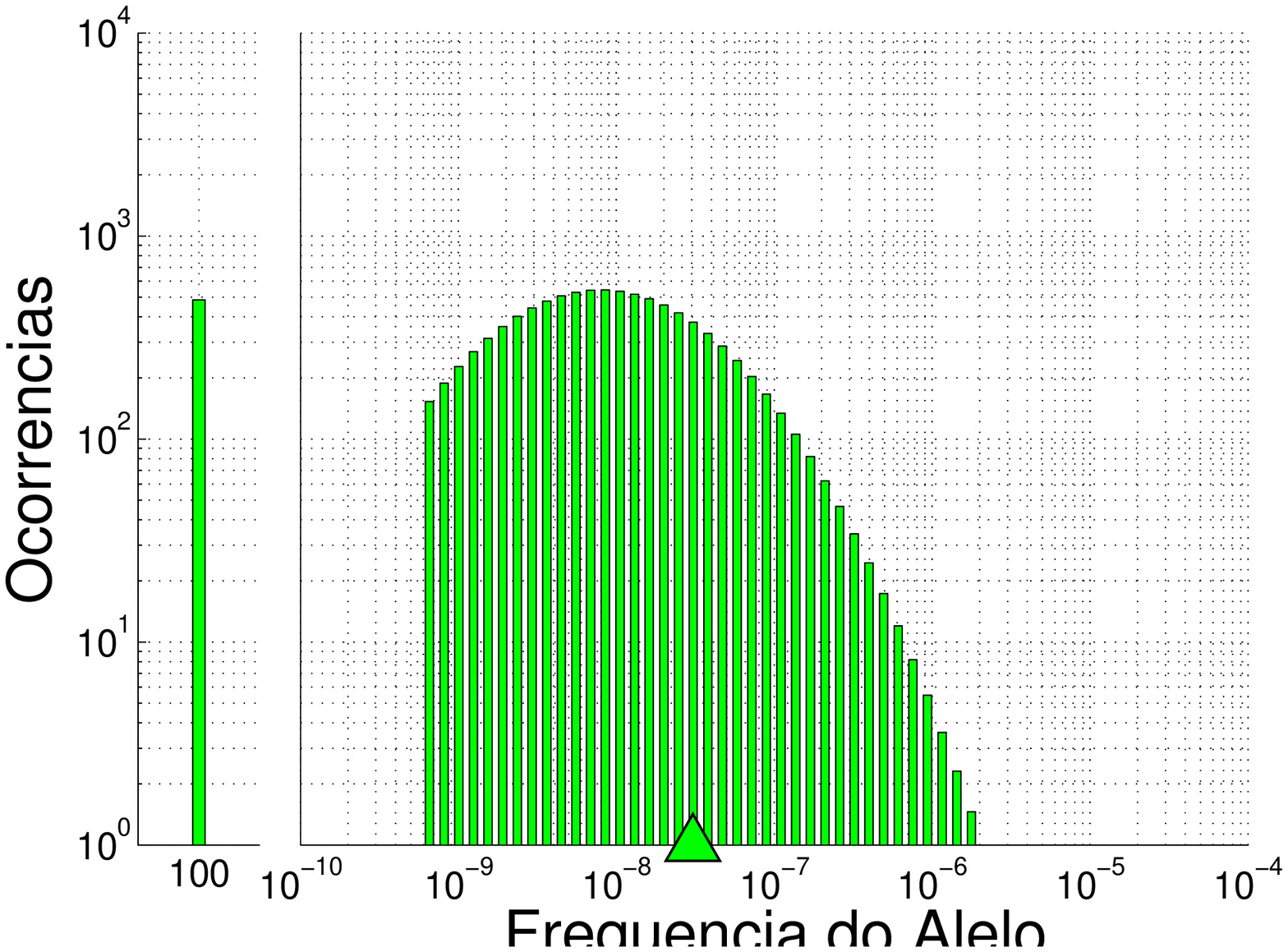}
\end{tabular*}
\caption{Allele frequency obtained by the stochastic model  at $t=0$ (on the left)
and by a log-normal distribution (on the right);  
histograms (green bar) and  averages (green triangle). 
The output of deterministic model is also displayed (red cross).}
\label{fig:FreqAllele2}
\end{figure}

\newpage
From the same simulation results we have estimated the histogram of
allele frequency at time $t=0$, as illustrated on the left of 
Figure \ref{fig:FreqAllele2}. 
Note that the number of occurrences of null resistant alleles is 
displayed aside, on the left of the plot.
This histogram give us an estimation for the \textit{wild distribution},  
and indicate a high probability of finding no resistant allele  
in the considered field. 
For comparison purposes, we also plot on the right of Figure \ref{fig:FreqAllele2}  
an histogram obtained from the log-normal distribution considered in 
\cite{Bagavathiannan_etal_2013,Neve_etal_2011,Neve_etal_2011B}. 
To make the comparison meaningful we keep the same mean obtained from the 
stochastic model (around $3.05\times 10^{-8}$)
and the standard deviation as in 
\cite{Bagavathiannan_etal_2013,Neve_etal_2011,Neve_etal_2011B}, set as $10^{-7}$,
and we also employ the same upper bound frequency 
given by the stochastic model
to calculate the null resistant alleles probability, that is, 
we set $P(\mu_0=0)=P(\mu_0 \leq 6.58\times 10^{-10})$ for the histogram of the 
log-normal distribution.


\subsection*{\bf Target herbicide  application period}
\label{sec:Res:compModels}

\begin{figure}
\begin{tabular*}{0.75\textwidth}{ c c }
\psfrag{10}[t][c]{\footnotesize 10} 
\psfrag{20}[t][c]{\footnotesize 20} 
\psfrag{30}[t][c]{\footnotesize 30} 
\psfrag{40}[t][c]{\footnotesize 40} 
\psfrag{50}[t][c]{\footnotesize 50} 
\psfrag{0}[b][l]{\footnotesize 0} 
\psfrag{0.2}[b][l]{\footnotesize 0.2} 
\psfrag{0.4}[b][l]{\footnotesize 0.4} 
\psfrag{0.6}[b][l]{\footnotesize 0.6} 
\psfrag{0.8}[b][l]{\footnotesize 0.8} 
\psfrag{1}[b][l]{\footnotesize 1}
\psfrag{Frequencia do Alelo}[b][c]{Allele frequency}
\psfrag{t}[t]{t}
\psfrag{LDXX}[bl]{\small $\mu(X_t)$}
\psfrag{LMXX}[bl]{\small $\mu({\mathfrak X}_t^{\omega_k})$}
\psfrag{LEXX}[bl]{\small $\mu({\bf X}_t^{\omega_k})$}
\includegraphics[height=5.1cm]{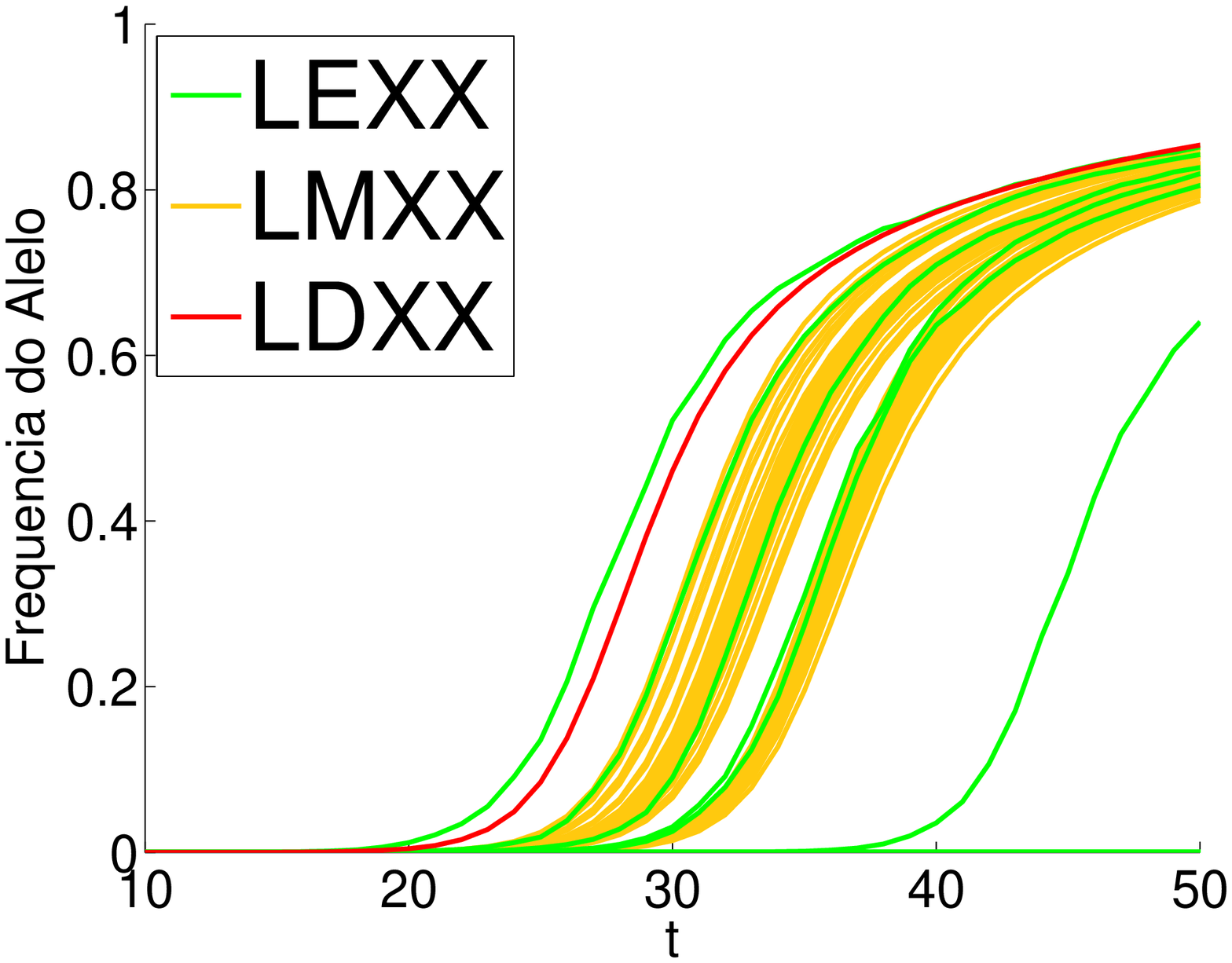} &
\psfrag{10}[t][c]{\footnotesize 10} 
\psfrag{20}[t][c]{\footnotesize 20} 
\psfrag{30}[t][c]{\footnotesize 30} 
\psfrag{40}[t][c]{\footnotesize 40} 
\psfrag{50}[t][c]{\footnotesize 50} 
\psfrag{0}[b][l]{\footnotesize 0} 
\psfrag{0.5}[b][l]{\footnotesize 0.5} 
\psfrag{1}[b][l]{\footnotesize 1} 
\psfrag{1.5}[b][l]{\footnotesize 1.5} 
\psfrag{2}[b][l]{\footnotesize 2}
\psfrag{2.5}[b][l]{\footnotesize 2.5}
\psfrag{3}[b][l]{\footnotesize 3}
\psfrag{x 10}{\scriptsize x 10}
\psfrag{4}[B][r][0.8]{\scriptsize 4}
\psfrag{Seeds.m-2}[b][c]{Seeds $m^{-2}$}
\psfrag{t}[t]{t}
\psfrag{LDXX}[bl]{\small $\eta(X_t)$}
\psfrag{LMXX}[bl]{\small $\eta({\mathfrak X}_t^{\omega_k})$}
\psfrag{LEXX}[bl]{\small $\eta({\bf X}_t^{\omega_k})$}
\includegraphics[height=5.1cm]{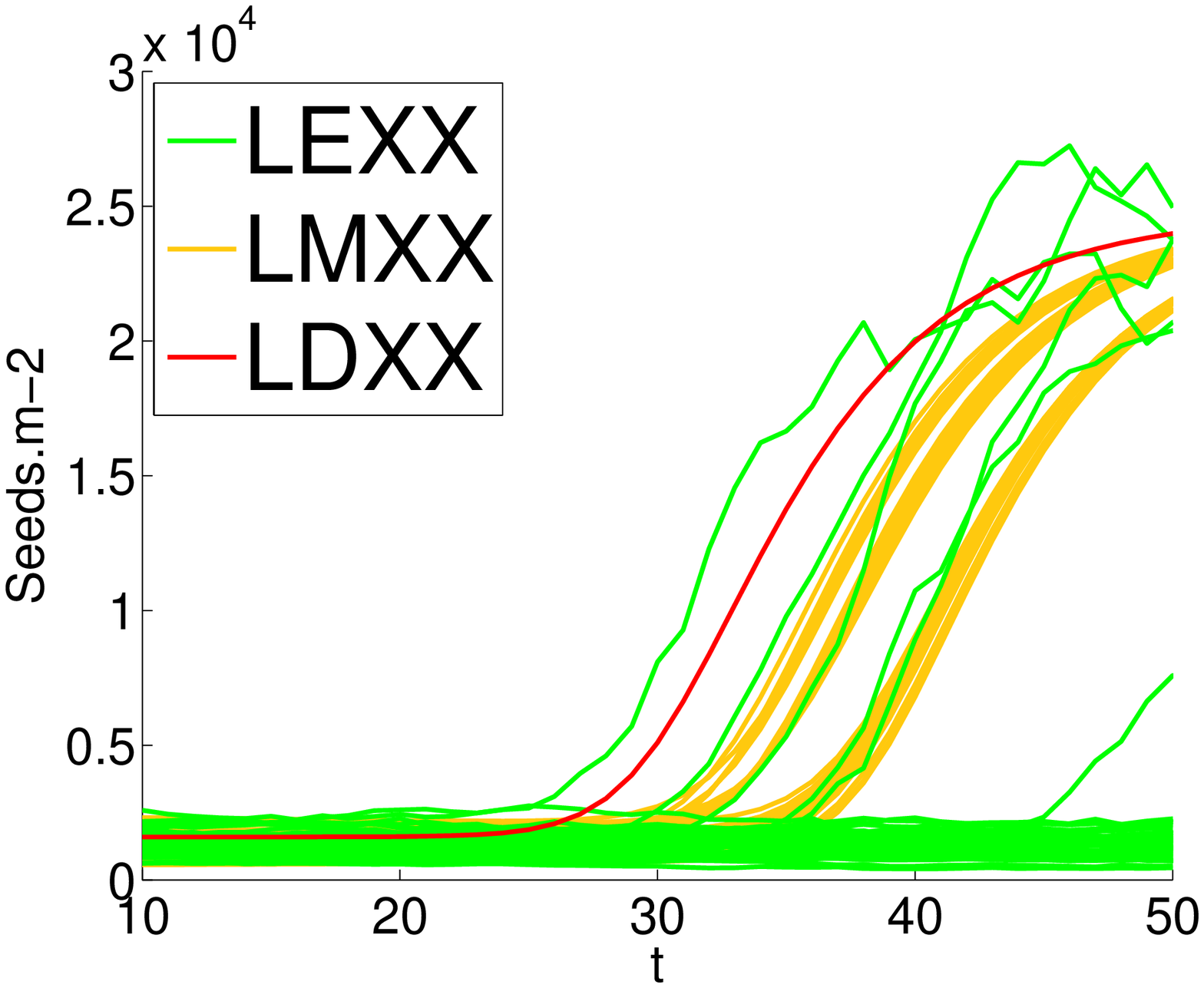} \\
\end{tabular*}
\caption{Allele frequency and seed bank trajectories given by the deterministic (red line), 
stochastic (green lines) and hybrid (yellow lines) models.}
\label{fig:FreqSeedsX}
\end{figure}

Figure \ref{fig:FreqSeedsX} shows some resistant allele frequency
and seed bank trajectories. 
In each figure we plot the deterministic trajectory and the stochastic 
and hybrid models trajectories associated to the realizations $k=1,\ldots 50$.
Every trajectory has an initial ``phase'' where it remains close to zero, and a second one 
in which it quickly gets far from zero,  referred as the 
\emph{resistance emergence period}. We say that this period starts at 
the \emph{emergence time}. Figure \ref{fig:FreqSeedsX} indicates 
that the emergence time is rather variable for the stochastic model, 
or more precisely, it is a random variable with large variance.
We note a smaller variance according to the hybrid model, 
and the simulations suggest that the emergence time is mostly constrained 
to a bounded time interval $t\in [20\; 30]$. 

\begin{center}
\begin{figure}

\begin{tabular*}{0.75\textwidth}{ c c }
\psfrag{100}[c][c][0.9]{\footnotesize $0$}
\psfrag{10}[c][l][0.9]{\footnotesize $10$}
\psfrag{1}[c][t][0.7]{\footnotesize $1$}
\psfrag{2}[c][t][0.7]{\footnotesize $2$}
\psfrag{3}[c][t][0.7]{\footnotesize $3$}
\psfrag{4}[c][t][0.7]{\footnotesize $4$}
\psfrag{-10}[tc][t][0.7]{\footnotesize $-10$}
\psfrag{-8}[tc][t][0.7]{\footnotesize $-8$}
\psfrag{-6}[tc][t][0.7]{\footnotesize $-6$}
\psfrag{-4}[tc][t][0.7]{\footnotesize $-4$}
\psfrag{-2}[tc][t][0.7]{\footnotesize $-2$}
\psfrag{0}[c][t][0.7]{\footnotesize $0$}
\psfrag{Ocorrencias}[b][c][1.2]{\footnotesize Occurrences}
\psfrag{Frequencia do Alelo}[b][b][1.2]{\footnotesize Allele Frequency}
\psfrag{LEXXX}[l][l][1.3]{\scriptsize $\overline{\mu}_{\bf X_1}$}
\psfrag{LDX}[l][l][1.3]{\scriptsize $\mu(X_1)$}
\includegraphics[scale=0.350]{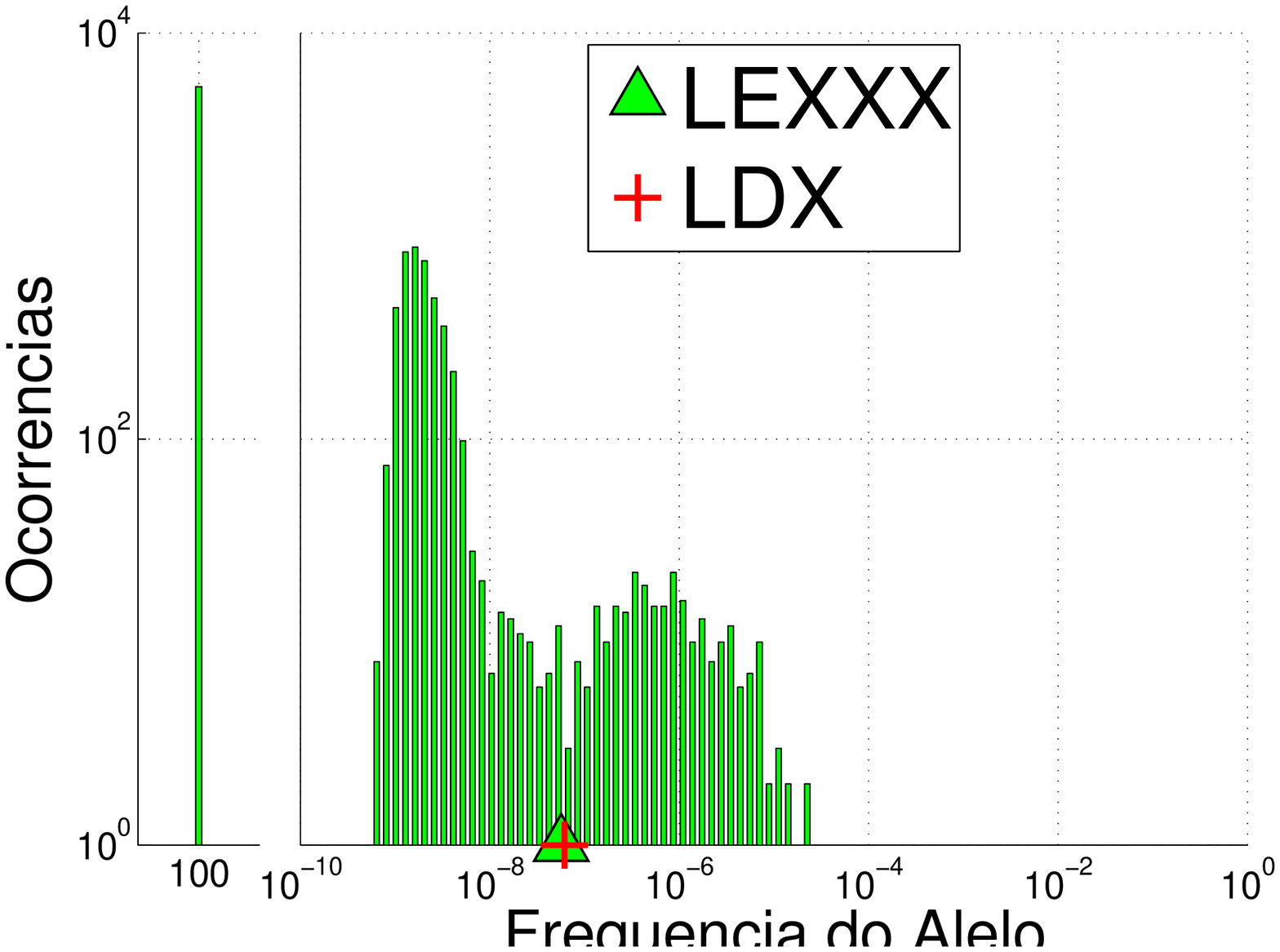}
 & 
\psfrag{100}[c][c][0.9]{\footnotesize $0$}
\psfrag{10}[c][l][0.9]{\footnotesize $10$}
\psfrag{1}[c][t][0.7]{\footnotesize $1$}
\psfrag{2}[c][t][0.7]{\footnotesize $2$}
\psfrag{3}[c][t][0.7]{\footnotesize $3$}
\psfrag{4}[c][t][0.7]{\footnotesize $4$}
\psfrag{-10}[tc][t][0.7]{\footnotesize $-10$}
\psfrag{-8}[tc][t][0.7]{\footnotesize $-8$}
\psfrag{-6}[tc][t][0.7]{\footnotesize $-6$}
\psfrag{-4}[tc][t][0.7]{\footnotesize $-4$}
\psfrag{-2}[tc][t][0.7]{\footnotesize $-2$}
\psfrag{0}[c][t][0.7]{\footnotesize $0$}
\psfrag{Ocorrencias}[b][c][1.2]{\footnotesize Occurrences}
\psfrag{Frequencia do Alelo}[b][b][1.2]{\footnotesize Allele Frequency}
\psfrag{LEXXX}[l][l][1.3]{\scriptsize $\overline{\mu}_{\bf X_5}$}
\psfrag{LDX}[l][l][1.3]{\scriptsize $\mu(X_5)$}
\includegraphics[scale=0.350]{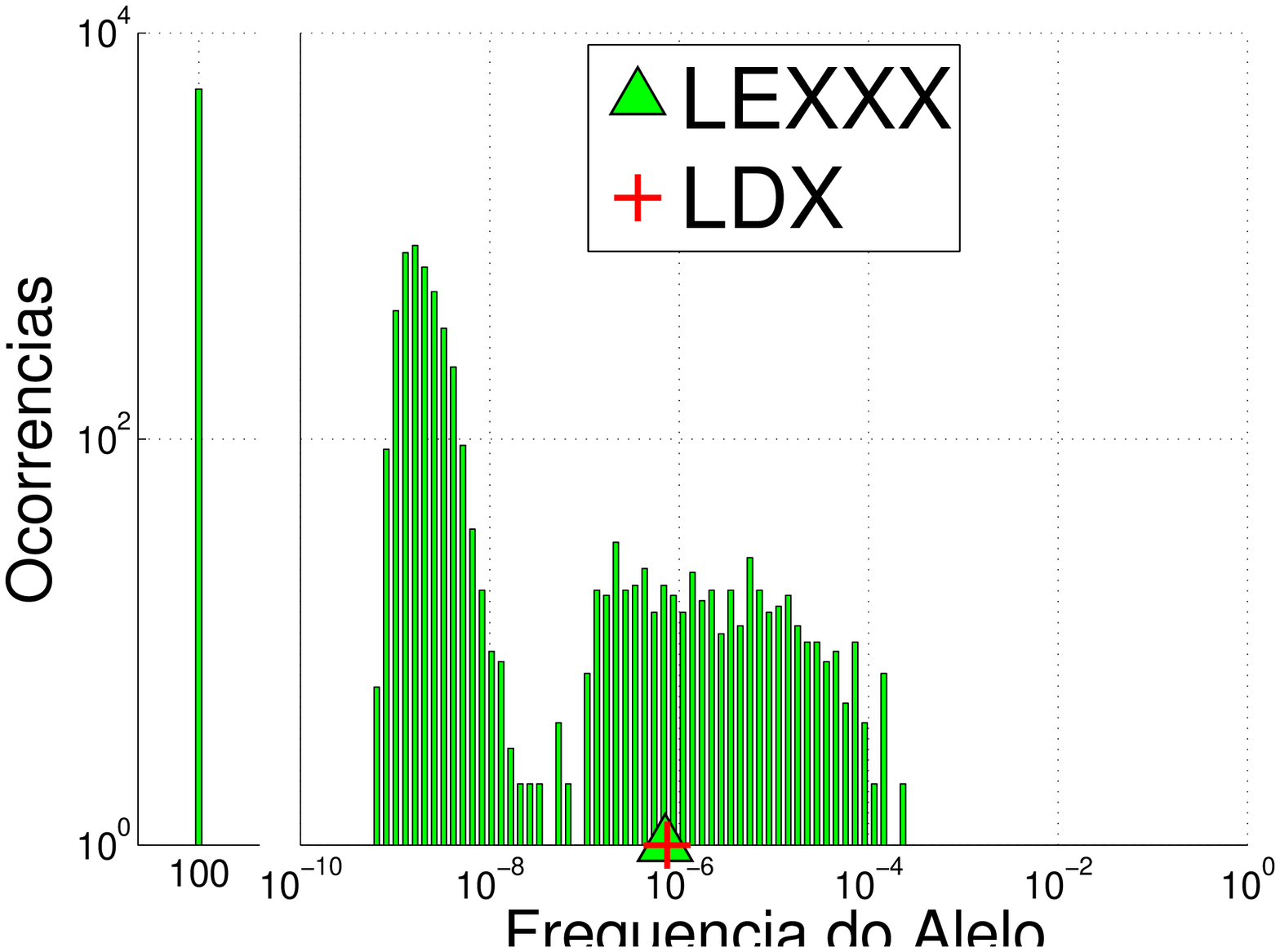} \\
\psfrag{100}[c][c][0.9]{\footnotesize $0$}
\psfrag{10}[c][l][0.9]{\footnotesize $10$}
\psfrag{1}[c][t][0.7]{\footnotesize $1$}
\psfrag{2}[c][t][0.7]{\footnotesize $2$}
\psfrag{3}[c][t][0.7]{\footnotesize $3$}
\psfrag{4}[c][t][0.7]{\footnotesize $4$}
\psfrag{-10}[tc][t][0.7]{\footnotesize $-10$}
\psfrag{-8}[tc][t][0.7]{\footnotesize $-8$}
\psfrag{-6}[tc][t][0.7]{\footnotesize $-6$}
\psfrag{-4}[tc][t][0.7]{\footnotesize $-4$}
\psfrag{-2}[tc][t][0.7]{\footnotesize $-2$}
\psfrag{0}[c][t][0.7]{\footnotesize $0$}
\psfrag{Ocorrencias}[b][c][1.2]{\footnotesize Occurrences}
\psfrag{Frequencia do Alelo}[b][b][1.2]{\footnotesize Allele Frequency}
\psfrag{LEXXX}[l][l][1.3]{\scriptsize $\overline{\mu}_{\bf X_{10}}$}
\psfrag{LDX}[l][l][1.3]{\scriptsize $\mu(X_{10})$}
\includegraphics[scale=0.350]{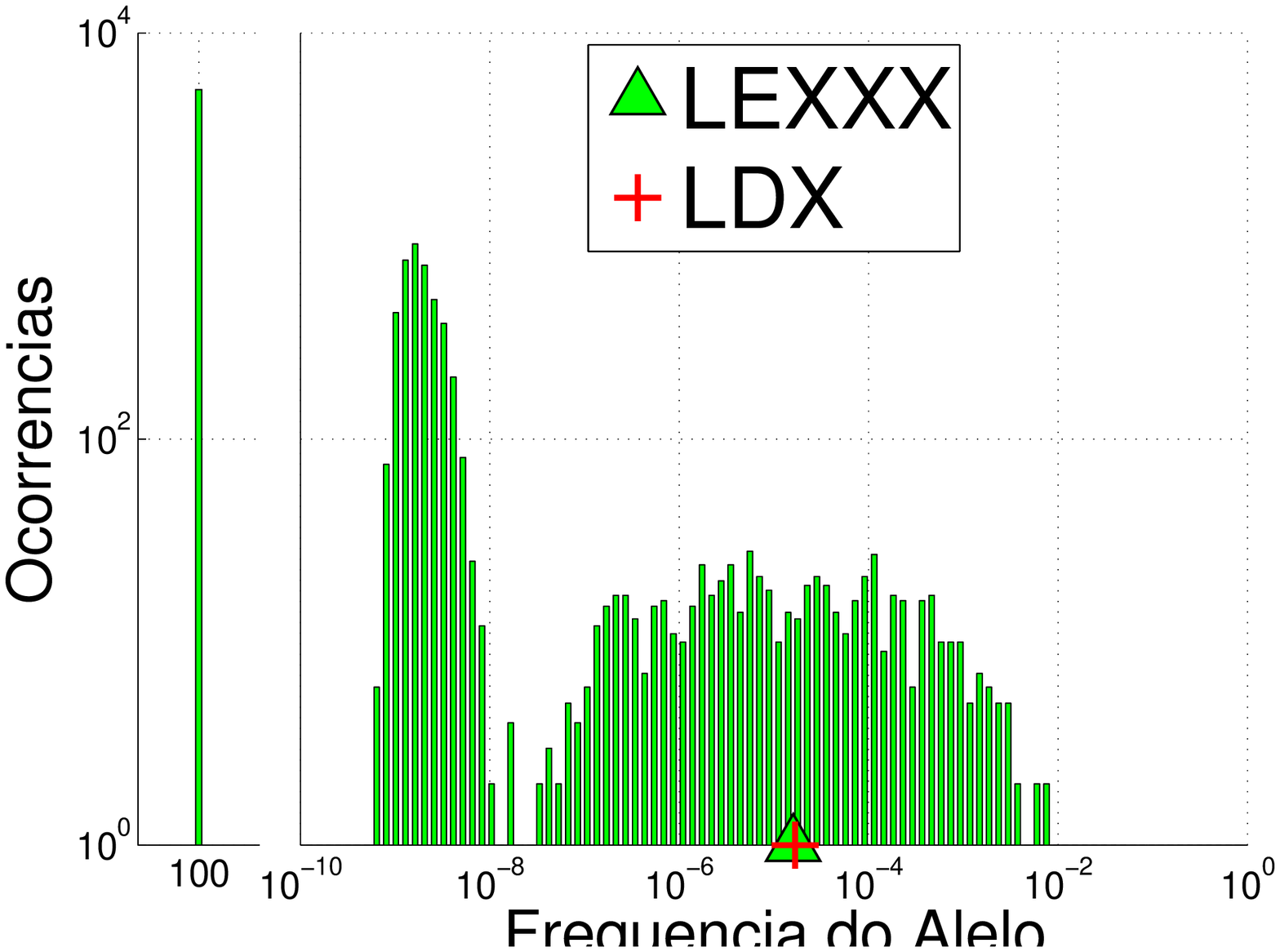}
&
\psfrag{100}[c][c][0.9]{\footnotesize $0$}
\psfrag{10}[c][l][0.9]{\footnotesize $10$}
\psfrag{1}[c][t][0.7]{\footnotesize $1$}
\psfrag{2}[c][t][0.7]{\footnotesize $2$}
\psfrag{3}[c][t][0.7]{\footnotesize $3$}
\psfrag{4}[c][t][0.7]{\footnotesize $4$}
\psfrag{-10}[tc][t][0.7]{\footnotesize $-10$}
\psfrag{-8}[tc][t][0.7]{\footnotesize $-8$}
\psfrag{-6}[tc][t][0.7]{\footnotesize $-6$}
\psfrag{-4}[tc][t][0.7]{\footnotesize $-4$}
\psfrag{-2}[tc][t][0.7]{\footnotesize $-2$}
\psfrag{0}[c][t][0.7]{\footnotesize $0$}
\psfrag{Ocorrencias}[b][c][1.2]{\footnotesize Occurrences}
\psfrag{Frequencia do Alelo}[b][b][1.2]{\footnotesize Allele Frequency}
\psfrag{LEXXX}[l][l][1.3]{\scriptsize $\overline{\mu}_{\bf X_{20}}$}
\psfrag{LDX}[l][l][1.3]{\scriptsize $\mu( X_{20})$}
\includegraphics[scale=0.350]{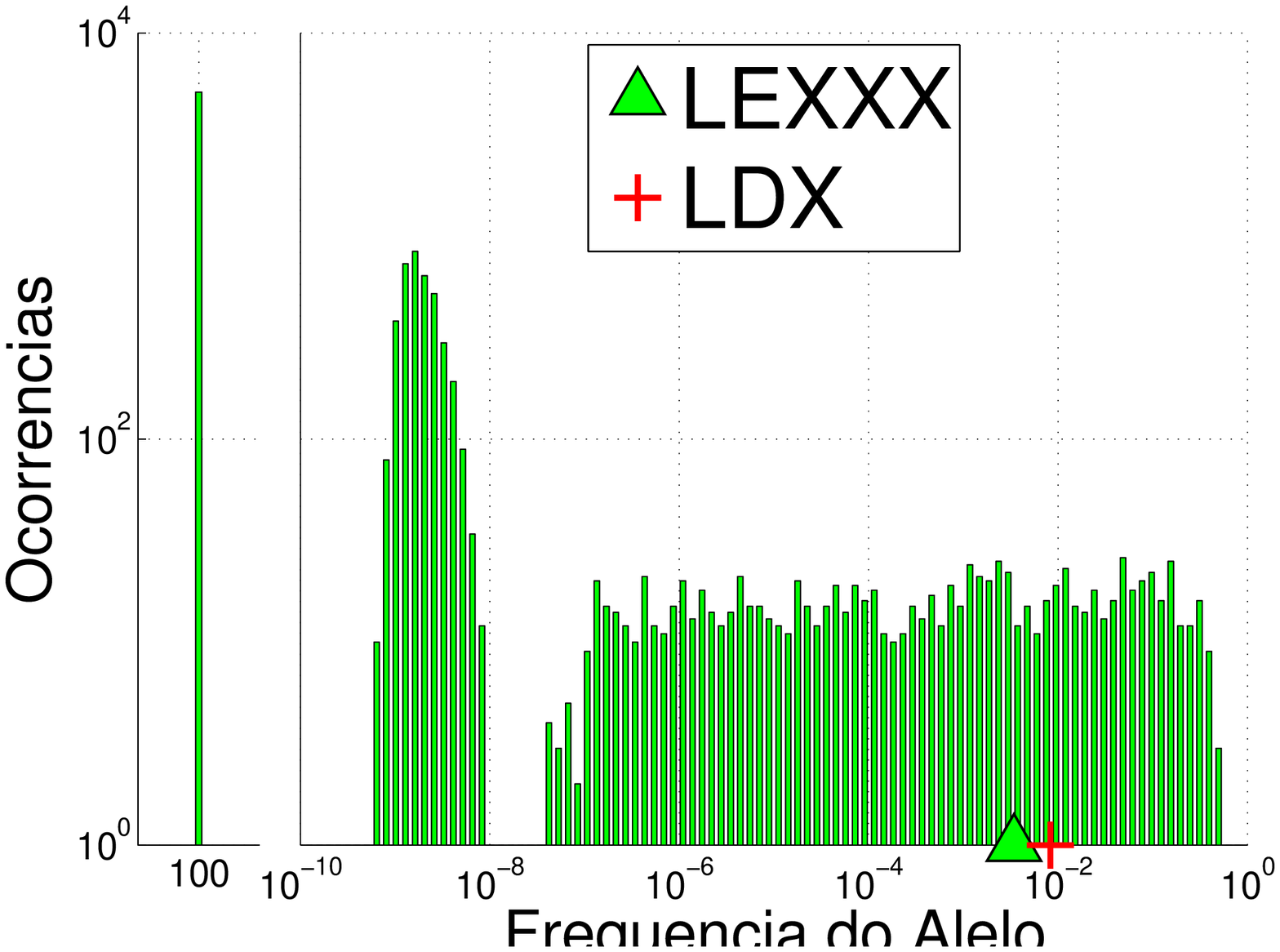}
\\
\psfrag{100}[c][c][0.9]{\footnotesize $0$}
\psfrag{10}[c][l][0.9]{\footnotesize $10$}
\psfrag{1}[c][t][0.7]{\footnotesize $1$}
\psfrag{2}[c][t][0.7]{\footnotesize $2$}
\psfrag{3}[c][t][0.7]{\footnotesize $3$}
\psfrag{4}[c][t][0.7]{\footnotesize $4$}
\psfrag{-10}[tc][t][0.7]{\footnotesize $-10$}
\psfrag{-8}[tc][t][0.7]{\footnotesize $-8$}
\psfrag{-6}[tc][t][0.7]{\footnotesize $-6$}
\psfrag{-4}[tc][t][0.7]{\footnotesize $-4$}
\psfrag{-2}[tc][t][0.7]{\footnotesize $-2$}
\psfrag{0}[c][t][0.7]{\footnotesize $0$}
\psfrag{Ocorrencias}[b][c][1.2]{\footnotesize Occurrences}
\psfrag{Frequencia do Alelo}[b][b][1.2]{\footnotesize Allele Frequency}
\psfrag{LEXXX}[l][l][1.3]{\scriptsize $\overline{\mu}_{\bf X_{50}}$}
\psfrag{LDX}[l][l][1.3]{\scriptsize $\mu(X_{50})$}
\includegraphics[scale=0.350]{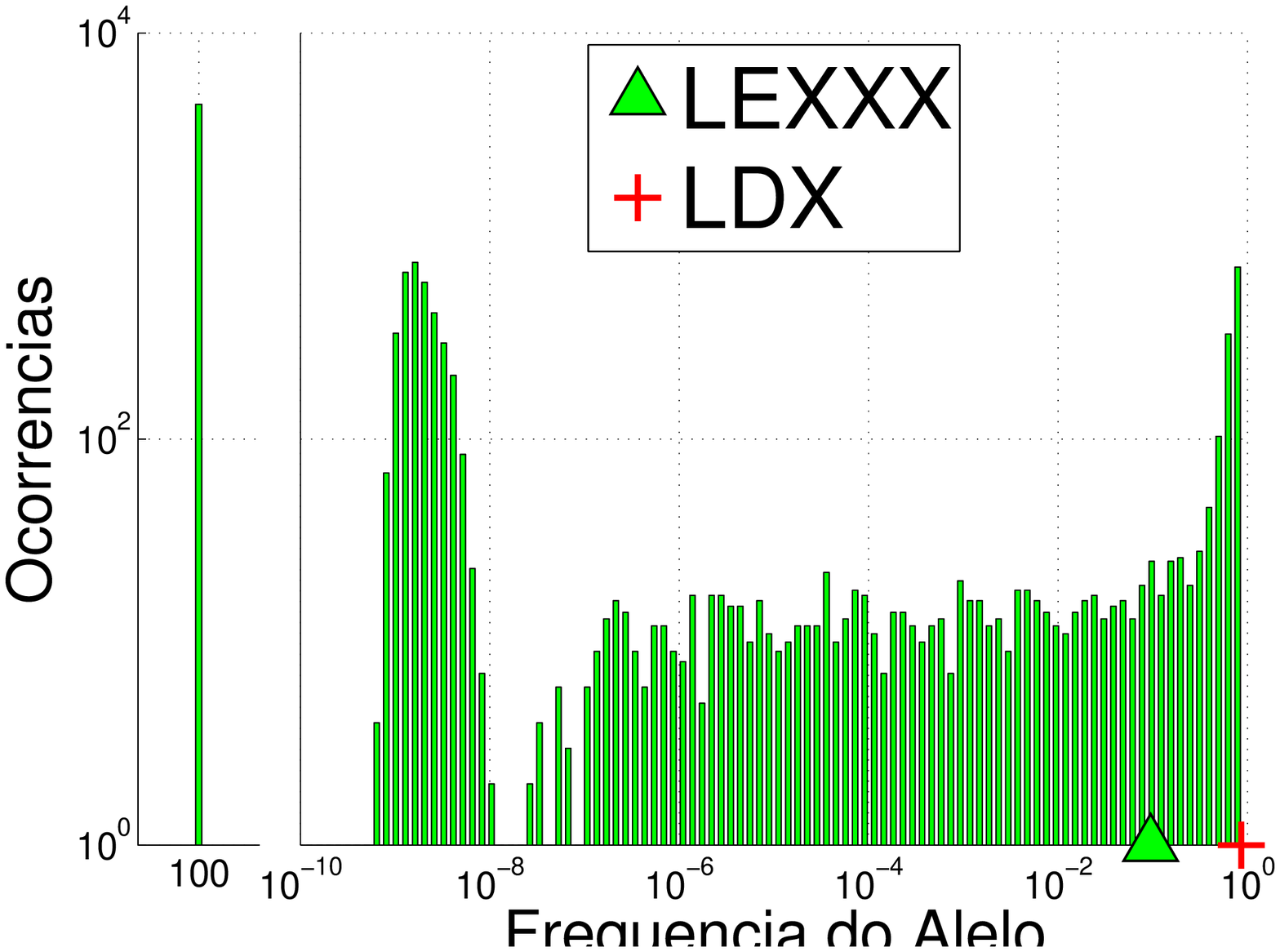}
& 
\psfrag{100}[c][c][0.9]{\footnotesize $0$}
\psfrag{10}[c][l][0.9]{\footnotesize $10$}
\psfrag{1}[c][t][0.7]{\footnotesize $1$}
\psfrag{2}[c][t][0.7]{\footnotesize $2$}
\psfrag{3}[c][t][0.7]{\footnotesize $3$}
\psfrag{4}[c][t][0.7]{\footnotesize $4$}
\psfrag{-10}[tc][t][0.7]{\footnotesize $-10$}
\psfrag{-8}[tc][t][0.7]{\footnotesize $-8$}
\psfrag{-6}[tc][t][0.7]{\footnotesize $-6$}
\psfrag{-4}[tc][t][0.7]{\footnotesize $-4$}
\psfrag{-2}[tc][t][0.7]{\footnotesize $-2$}
\psfrag{0}[c][t][0.7]{\footnotesize $0$}
\psfrag{Ocorrencias}[b][c][1.2]{\footnotesize Occurrences}
\psfrag{Frequencia do Alelo}[b][b][1.2]{\footnotesize Allele Frequency}
\psfrag{LEXXX}[l][l][1.3]{\scriptsize $\overline{\mu}_{\bf X_{100}}$}
\psfrag{LDX}[l][l][1.3]{\scriptsize $\mu(X_{100})$}
\includegraphics[scale=0.350]{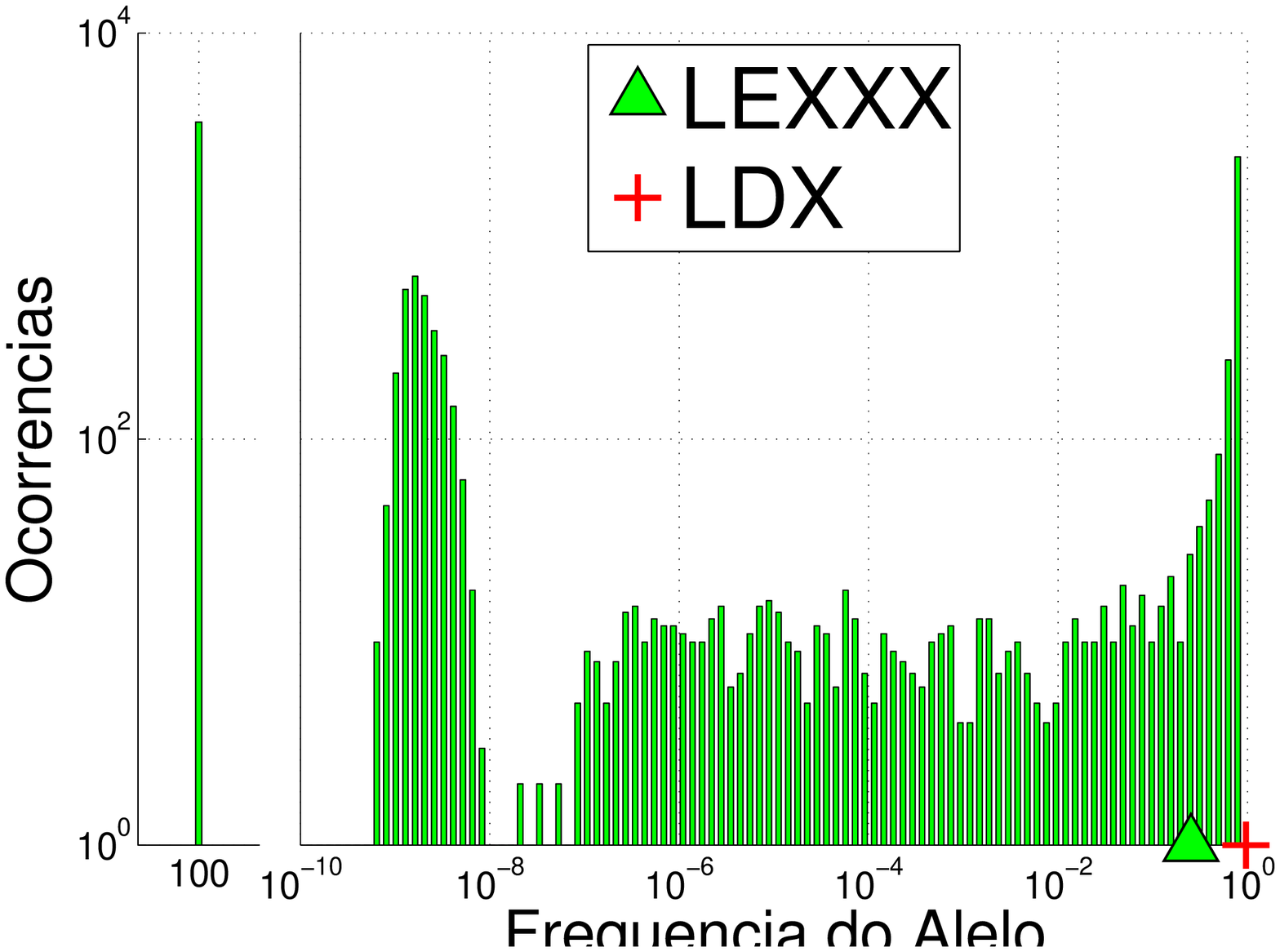}
\end{tabular*}

\caption{Histogram of allele frequency obtained from the stochastic  model 
for $t = 1,5,10,20,50,100$.
The green triangles indicate the average of distributions.
The red crosses give the  values obtained from the deterministic model.}
\label{fig:FreqAlleleAaplicSTOSTO1}
\end{figure}
\end{center}

\begin{center}
\begin{figure}

\begin{tabular*}{0.75\textwidth}{ c c }
\psfrag{100}[c][c][0.9]{\footnotesize $0$}
\psfrag{10}[c][l][0.9]{\footnotesize $10$}
\psfrag{1}[c][t][0.7]{\footnotesize $1$}
\psfrag{2}[c][t][0.7]{\footnotesize $2$}
\psfrag{3}[c][t][0.7]{\footnotesize $3$}
\psfrag{4}[c][t][0.7]{\footnotesize $4$}
\psfrag{-10}[tc][t][0.7]{\footnotesize $-10$}
\psfrag{-8}[tc][t][0.7]{\footnotesize $-8$}
\psfrag{-6}[tc][t][0.7]{\footnotesize $-6$}
\psfrag{-4}[tc][t][0.7]{\footnotesize $-4$}
\psfrag{-2}[tc][t][0.7]{\footnotesize $-2$}
\psfrag{0}[c][t][0.7]{\footnotesize $0$}
\psfrag{Ocorrencias}[b][c][1.2]{\footnotesize Occurrences}
\psfrag{Frequencia do Alelo}[b][b][1.2]{\footnotesize Allele Frequency}
\psfrag{LMXXX}[l][l][1.3]{\scriptsize $\overline{\mu}_{\mathfrak X_1}$}
\psfrag{LDX}[l][l][1.3]{\scriptsize $\mu( X_1)$}
\includegraphics[scale=0.350]{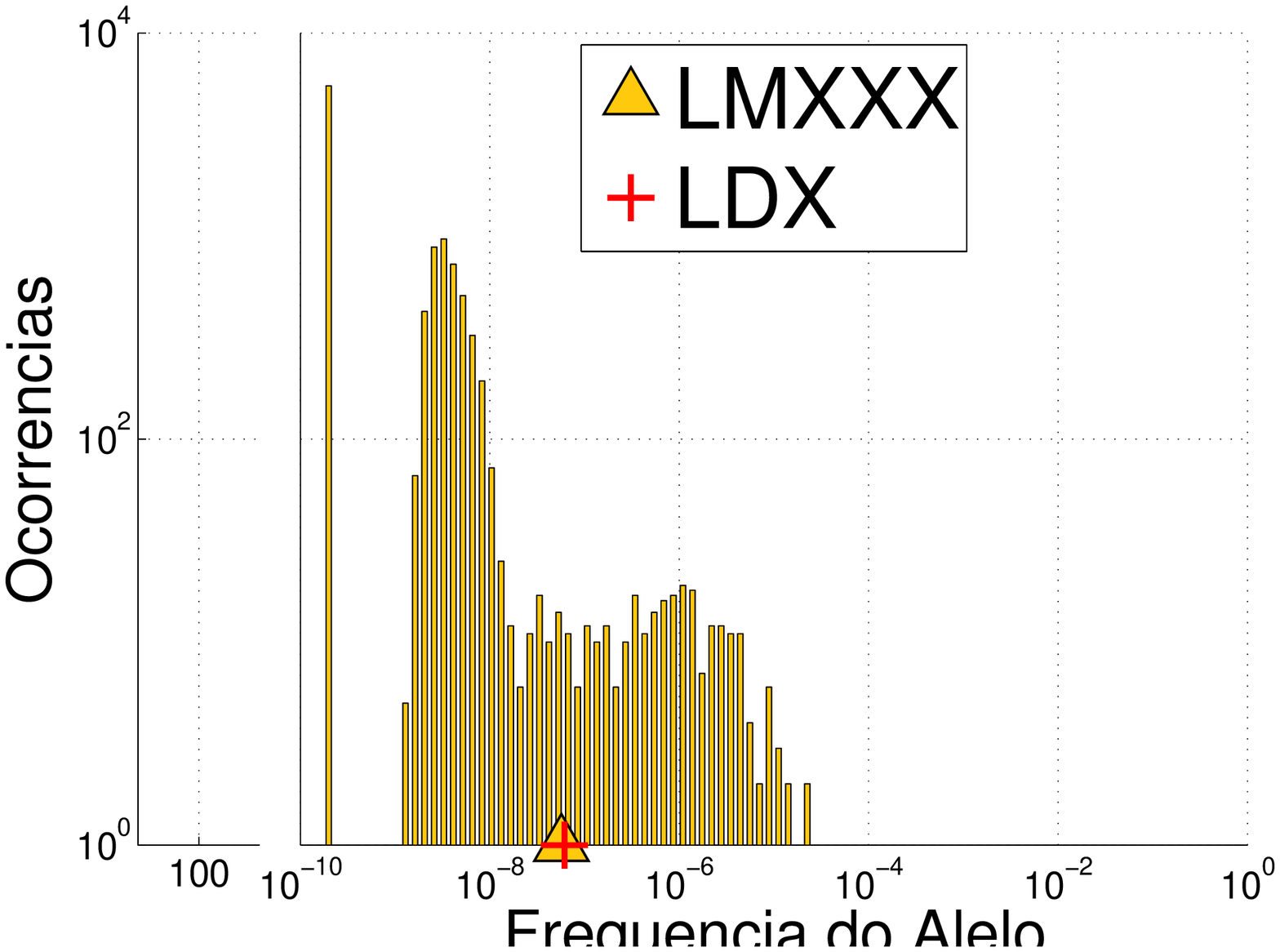}
&
\psfrag{100}[c][c][0.9]{\footnotesize $0$}
\psfrag{10}[c][l][0.9]{\footnotesize $10$}
\psfrag{1}[c][t][0.7]{\footnotesize $1$}
\psfrag{2}[c][t][0.7]{\footnotesize $2$}
\psfrag{3}[c][t][0.7]{\footnotesize $3$}
\psfrag{4}[c][t][0.7]{\footnotesize $4$}
\psfrag{-10}[tc][t][0.7]{\footnotesize $-10$}
\psfrag{-8}[tc][t][0.7]{\footnotesize $-8$}
\psfrag{-6}[tc][t][0.7]{\footnotesize $-6$}
\psfrag{-4}[tc][t][0.7]{\footnotesize $-4$}
\psfrag{-2}[tc][t][0.7]{\footnotesize $-2$}
\psfrag{0}[c][t][0.7]{\footnotesize $0$}
\psfrag{Ocorrencias}[b][c][1.2]{\footnotesize Occurrences}
\psfrag{Frequencia do Alelo}[b][b][1.2]{\footnotesize Allele Frequency}
\psfrag{LMXXX}[l][l][1.3]{\scriptsize $\overline{\mu}_{\mathfrak X_5}$}
\psfrag{LDX}[l][l][1.3]{\scriptsize $\mu( X_5)$}
\includegraphics[scale=0.350]{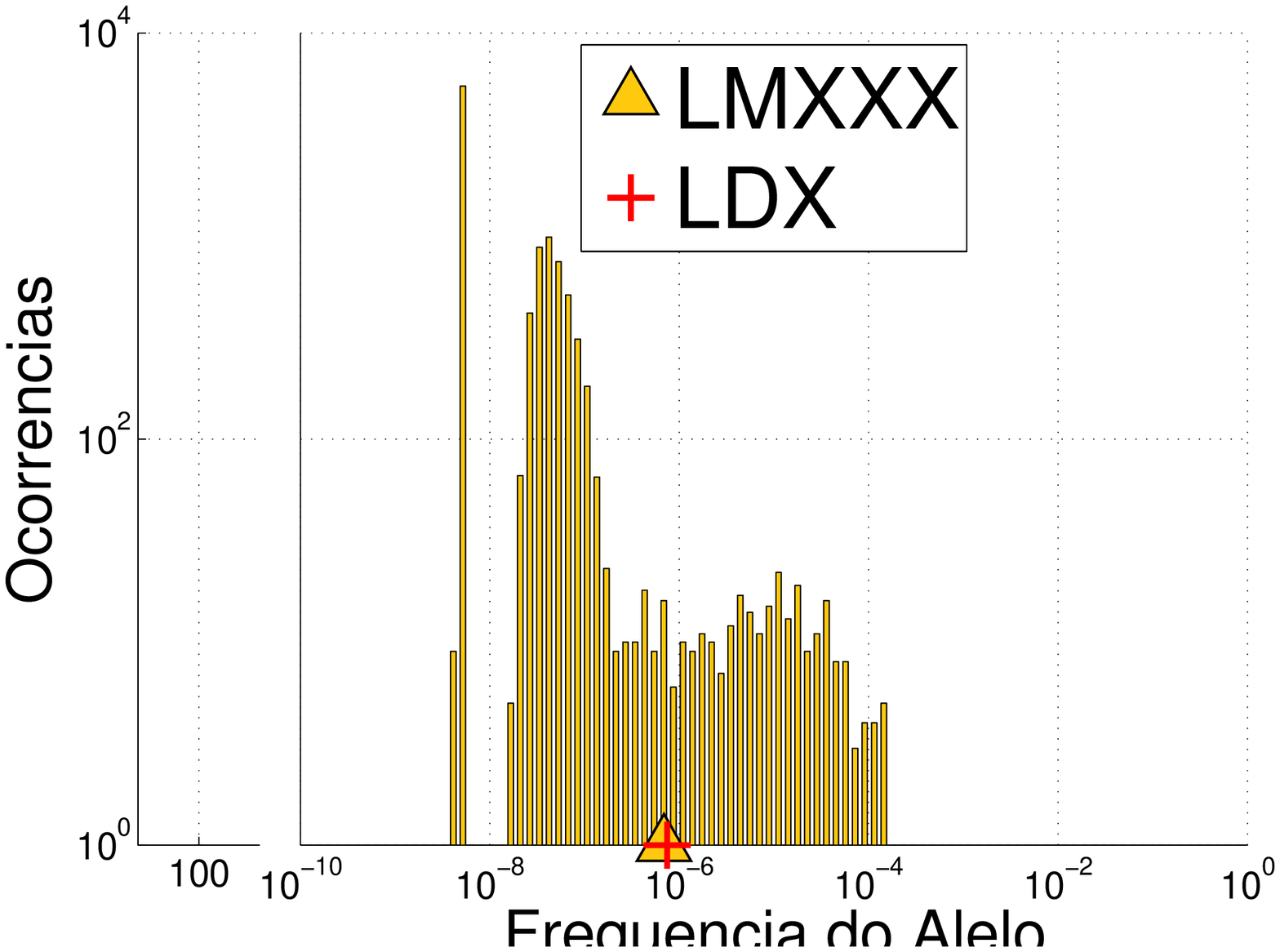} \\
\psfrag{100}[c][c][0.9]{\footnotesize $0$}
\psfrag{10}[c][l][0.9]{\footnotesize $10$}
\psfrag{1}[c][t][0.7]{\footnotesize $1$}
\psfrag{2}[c][t][0.7]{\footnotesize $2$}
\psfrag{3}[c][t][0.7]{\footnotesize $3$}
\psfrag{4}[c][t][0.7]{\footnotesize $4$}
\psfrag{-10}[tc][t][0.7]{\footnotesize $-10$}
\psfrag{-8}[tc][t][0.7]{\footnotesize $-8$}
\psfrag{-6}[tc][t][0.7]{\footnotesize $-6$}
\psfrag{-4}[tc][t][0.7]{\footnotesize $-4$}
\psfrag{-2}[tc][t][0.7]{\footnotesize $-2$}
\psfrag{0}[c][t][0.7]{\footnotesize $0$}
\psfrag{Ocorrencias}[b][c][1.2]{\footnotesize Occurrences}
\psfrag{Frequencia do Alelo}[b][b][1.2]{\footnotesize Allele Frequency}
\psfrag{LMXXX}[l][l][1.3]{\scriptsize $\overline{\mu}_{\mathfrak X_{10}}$}
\psfrag{LDX}[l][l][1.3]{\scriptsize $\mu( X_{10})$}
\includegraphics[scale=0.350]{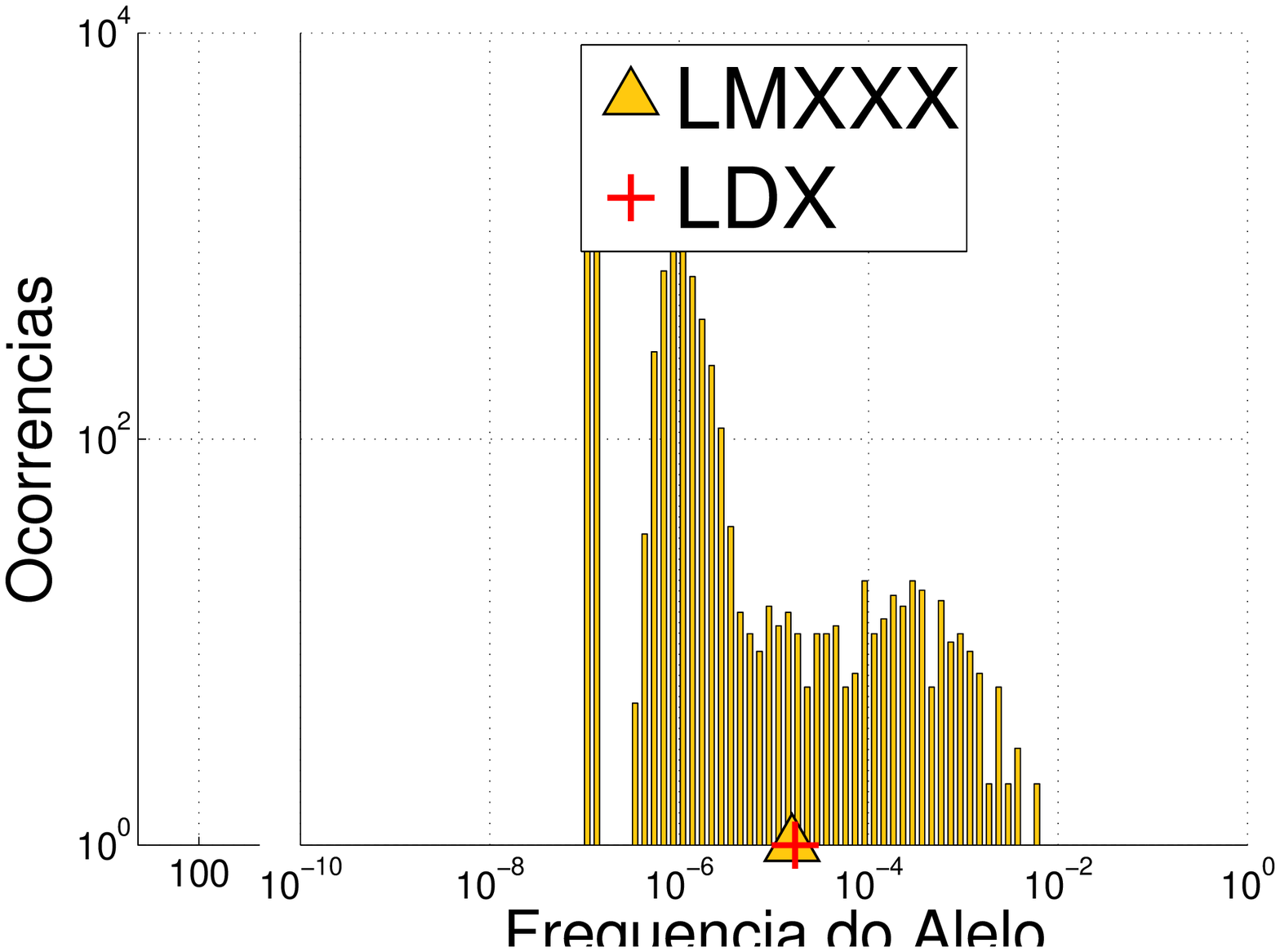}
&
\psfrag{100}[c][c][0.9]{\footnotesize $0$}
\psfrag{10}[c][l][0.9]{\footnotesize $10$}
\psfrag{1}[c][t][0.7]{\footnotesize $1$}
\psfrag{2}[c][t][0.7]{\footnotesize $2$}
\psfrag{3}[c][t][0.7]{\footnotesize $3$}
\psfrag{4}[c][t][0.7]{\footnotesize $4$}
\psfrag{-10}[tc][t][0.7]{\footnotesize $-10$}
\psfrag{-8}[tc][t][0.7]{\footnotesize $-8$}
\psfrag{-6}[tc][t][0.7]{\footnotesize $-6$}
\psfrag{-4}[tc][t][0.7]{\footnotesize $-4$}
\psfrag{-2}[tc][t][0.7]{\footnotesize $-2$}
\psfrag{0}[c][t][0.7]{\footnotesize $0$}
\psfrag{Ocorrencias}[b][c][1.2]{\footnotesize Occurrences}
\psfrag{Frequencia do Alelo}[b][b][1.2]{\footnotesize Allele Frequency}
\psfrag{LMXXX}[l][l][1.3]{\scriptsize $\overline{\mu}_{\mathfrak X_{20}}$}
\psfrag{LDX}[l][l][1.3]{\scriptsize $\mu( X_{20})$}
\includegraphics[scale=0.350]{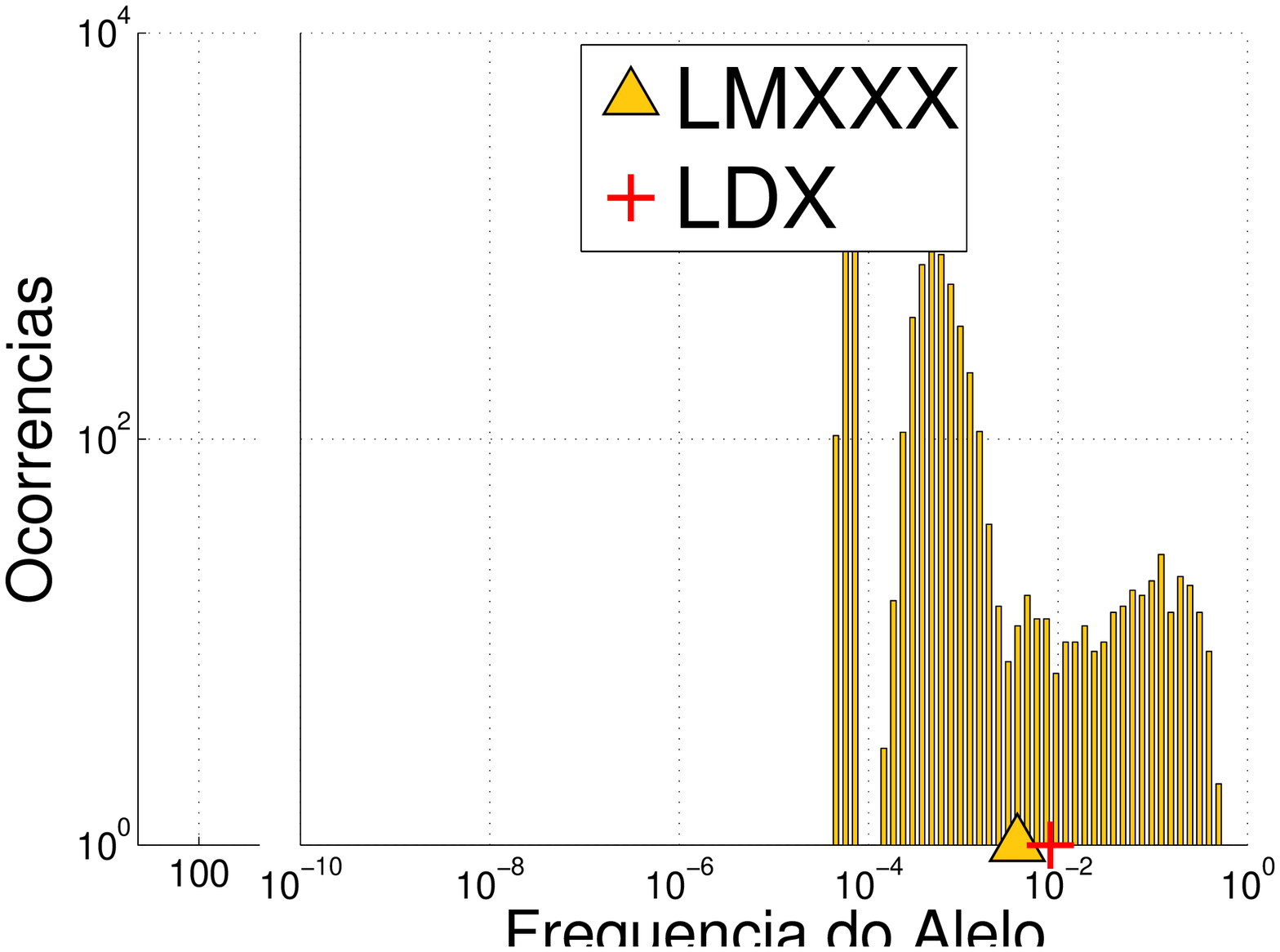} \\
\psfrag{100}[c][c][0.9]{\footnotesize $0$}
\psfrag{10}[c][l][0.9]{\footnotesize $10$}
\psfrag{1}[c][t][0.7]{\footnotesize $1$}
\psfrag{2}[c][t][0.7]{\footnotesize $2$}
\psfrag{3}[c][t][0.7]{\footnotesize $3$}
\psfrag{4}[c][t][0.7]{\footnotesize $4$}
\psfrag{-10}[tc][t][0.7]{\footnotesize $-10$}
\psfrag{-8}[tc][t][0.7]{\footnotesize $-8$}
\psfrag{-6}[tc][t][0.7]{\footnotesize $-6$}
\psfrag{-4}[tc][t][0.7]{\footnotesize $-4$}
\psfrag{-2}[tc][t][0.7]{\footnotesize $-2$}
\psfrag{0}[c][t][0.7]{\footnotesize $0$}
\psfrag{Ocorrencias}[b][c][1.2]{\footnotesize Occurrences}
\psfrag{Frequencia do Alelo}[b][b][1.2]{\footnotesize Allele Frequency}
\psfrag{LMXXX}[l][l][1.3]{\scriptsize $\overline{\mu}_{\mathfrak X_{50}}$}
\psfrag{LDX}[l][l][1.3]{\scriptsize $\mu( X_{50})$}
\includegraphics[scale=0.350]{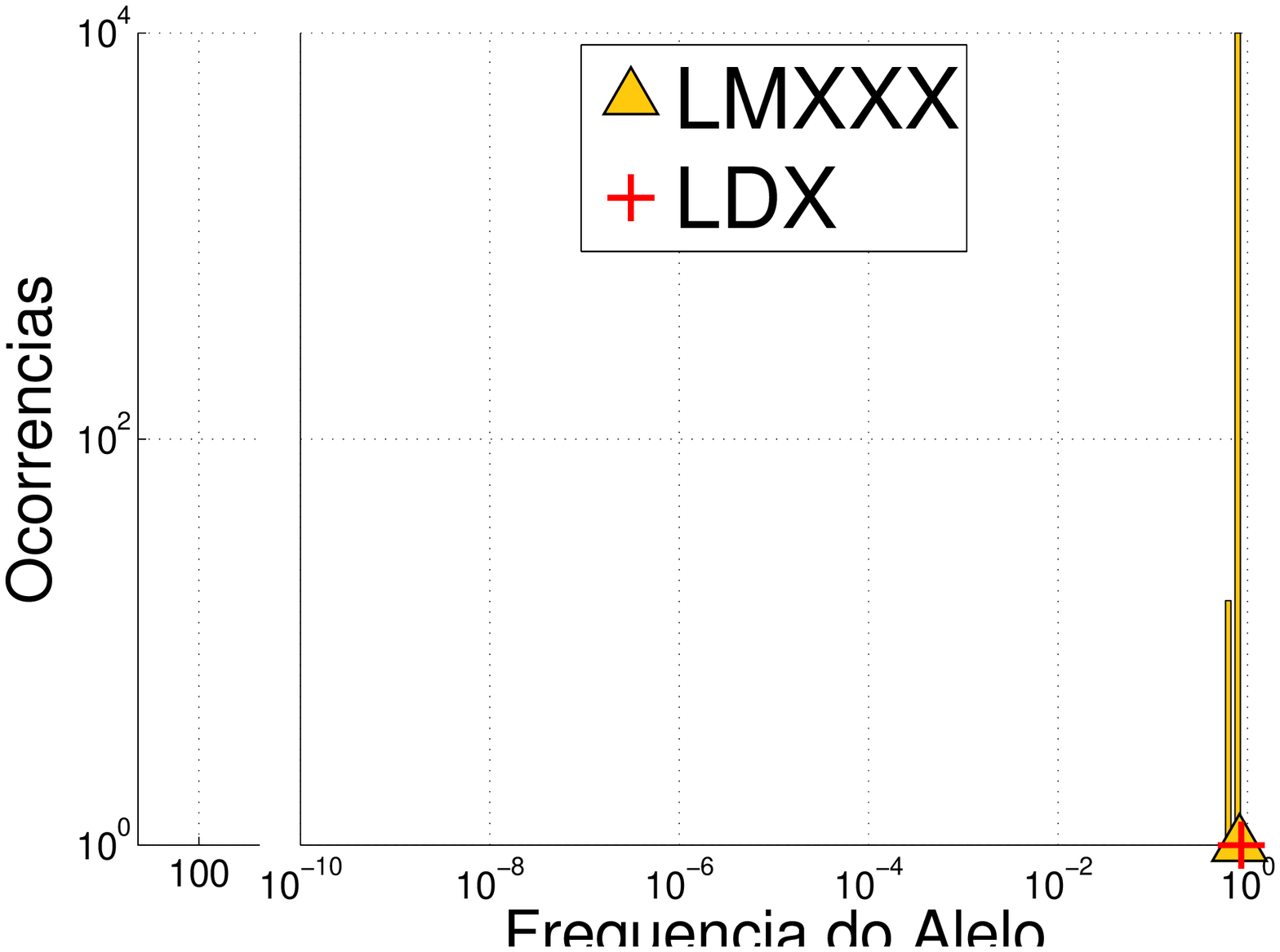}
&
\psfrag{100}[c][c][0.9]{\footnotesize $0$}
\psfrag{10}[c][l][0.9]{\footnotesize $10$}
\psfrag{1}[c][t][0.7]{\footnotesize $1$}
\psfrag{2}[c][t][0.7]{\footnotesize $2$}
\psfrag{3}[c][t][0.7]{\footnotesize $3$}
\psfrag{4}[c][t][0.7]{\footnotesize $4$}
\psfrag{-10}[tc][t][0.7]{\footnotesize $-10$}
\psfrag{-8}[tc][t][0.7]{\footnotesize $-8$}
\psfrag{-6}[tc][t][0.7]{\footnotesize $-6$}
\psfrag{-4}[tc][t][0.7]{\footnotesize $-4$}
\psfrag{-2}[tc][t][0.7]{\footnotesize $-2$}
\psfrag{0}[c][t][0.7]{\footnotesize $0$}
\psfrag{Ocorrencias}[b][c][1.2]{\footnotesize Occurrences}
\psfrag{Frequencia do Alelo}[b][b][1.2]{\footnotesize Allele Frequency}
\psfrag{LMXXX}[l][l][1.3]{\scriptsize $\overline{\mu}_{\mathfrak X_{100}}$}
\psfrag{LDX}[l][l][1.3]{\scriptsize $\mu( X_{100})$}
\includegraphics[scale=0.350]{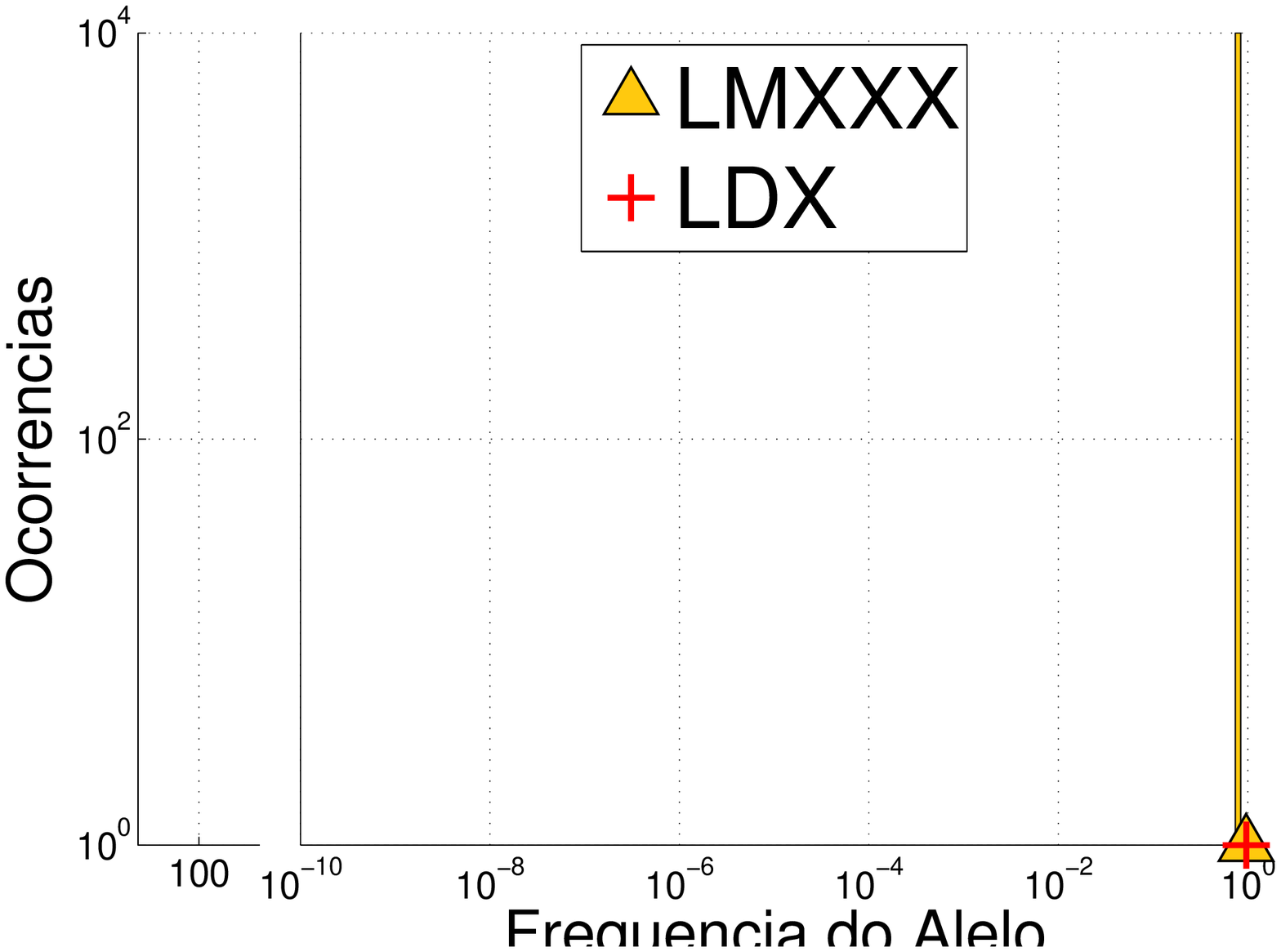}
\end{tabular*}

\caption{Histogram of allele frequency obtained from the hybrid  model 
for $t = 1,5,10,20,50,100$.
The yellow triangles indicate the average of distributions.
The red crosses give the  values obtained from the deterministic model.}
\label{fig:FreqAlleleAaplicSTOSTO2}
\end{figure}
\end{center}

Distributions of the resistant allele frequencies estimated by the
stochastic and hybrid models for $t = 1,5,10,20,50,100$ are shown in Figures
\ref{fig:FreqAlleleAaplicSTOSTO1} and
\ref{fig:FreqAlleleAaplicSTOSTO2}, respectively.
These figures illustrate how both models predict the evolution of 
the wild distribution (Figure \ref{fig:FreqAllele2}) 
along the time. 
It is interesting to note that for $t\geq 5$, every histogram of the
stochastic model (Figure \ref{fig:FreqAlleleAaplicSTOSTO1}) 
shows a portion of low occurrences around the interval
$[10^{-8} \; 10^{-7}]$, which we refer to as the 
\emph{few occurrences region (FOR)}.  
We also note to the left of the FOR,
around $[10^{-9} \; 10^{-8}]$, a region with relatively many
occurrences, which we call \emph{left peak} for short.  
We believe that the left peaks are maintained almost
exclusively by the mutation phenomenon (a $\rightarrow$ A), 
and not by seeds generated by
the reproduction of resistant mature plants.  
Note that the bar corresponding to the null resistant allele occurrences remains high even
after one hundred cycles of herbicide application.
 Note also that for $t\geq 20$ the deterministic model outputs 
differs considerably from the expected value calculated from the 
stochastic model. Regarding the hybrid model 
(Figure \ref{fig:FreqAlleleAaplicSTOSTO2}) we see how the deterministic
dynamics after $t=0$ modifies the wild distribution: it constantly shifts
the distribution to the right towards the bound $\mu = 10^0$.

\par

\begin{figure}[!ht]
\psfrag{0}{\scriptsize 0} 
\psfrag{1}{\scriptsize 1} 
\psfrag{2}{\scriptsize 2} 
\psfrag{3}{\scriptsize 3} 
\psfrag{4}{\scriptsize 4} 
\psfrag{10}{\footnotesize 10} 
\psfrag{15}{\footnotesize 15} 
\psfrag{20}{\footnotesize 20} 
\psfrag{25}{\footnotesize 25} 
\psfrag{30}{\footnotesize 30} 
\psfrag{35}{\footnotesize 35} 
\psfrag{40}{\footnotesize 40} 
\psfrag{50}{\footnotesize 50} 
\psfrag{2000}{\footnotesize 2000} 
\psfrag{4000}{\footnotesize 4000} 
\psfrag{6000}{\footnotesize 6000} 
\psfrag{8000}{\footnotesize 8000} 
\psfrag{10000}{\footnotesize 10000} 
\psfrag{E}[bl]{\small $f^t_{\bf X}$}
\psfrag{DE}[bl]{\small $f^t_{\mathfrak X}$}
\psfrag{D}[bl]{\small $f^t_{ X}$}
\psfrag{Ocorrencias}{Occurrences}
\psfrag{t}{t}
 \centering
\includegraphics[scale=0.45]{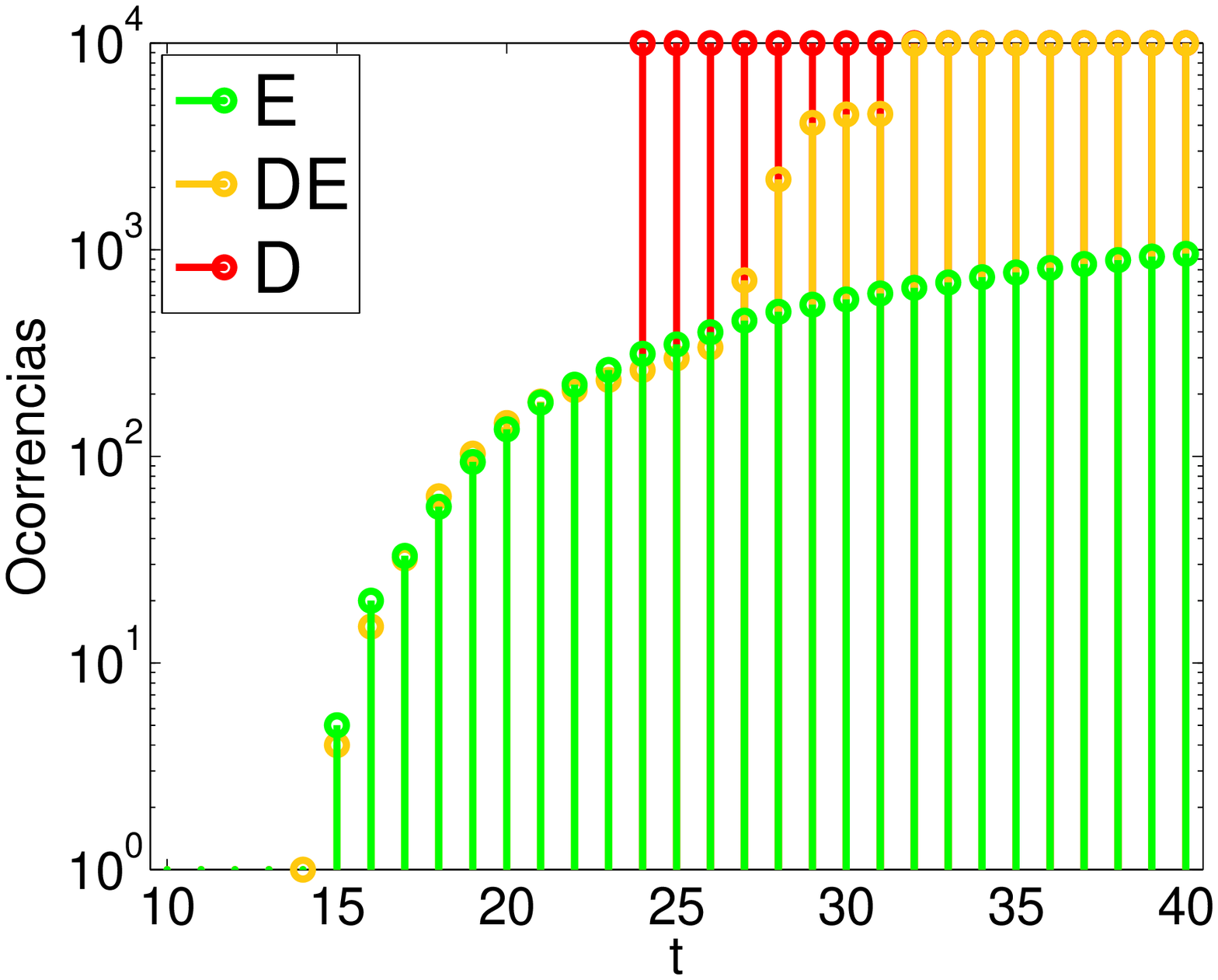}
\caption{Accumulated occurrence of weed resistance visualization 
predicted by the models.}
\label{fig:IdentRes}
\end{figure}

We have obtained the resistance visualization 
from the simulated trajectories 
${\bf X}_t$, $\mathfrak{X}_t$ and $X_t$ according to \eqref{eq:indexResis}. 
The accumulated number of occurrences of weed visualization 
\eqref{eq:identResFig} 
illustrated in Figure \ref{fig:IdentRes} 
shows a booming visualization at time $t=14$ and 
similar results in the interval $15\leq t\leq 26$ using the hybrid and 
stochastic models. From $t=27$ on the occurrence estimated by the hybrid 
model increases considerably
faster than the one given by the stochastic model, such that at $t = 32$ 
we have $f_{\mathfrak{X}}^t=10^4$, i.e., we have visualization of resistance 
in all realizations. 
The deterministic model exhibits a rather simple behaviour: 
there is no visualization 
of resistance prior to $t=24$, and $100\%$ of visualization at $t=24$.

\section{Discussion}
\label{sec-discussion}

The stochastic model takes advantage of the inherent stability of the 
dynamics to weaken the dependence of the simulation results for $t\geq 0$ 
on the initial guess for the allele frequency at time $-T_0$; 
this allowed us to use the strategy of setting $-T_0$ in a distant past
and to overcome the difficulty in finding a meaningful distribution for $\mu_0$.
It is unfruitful, for instance, to  use the same strategy for the deterministic 
model, as the distribution of $\mu_0$ turns out to be very concentrated
around the value given by the Mutation-Selection Theory.

The distribution of $\mu_0$ obtained with our model and the 
log-normal distribution given in Figure \ref{fig:FreqXSeedDensity} 
disagree. Of particular relevance is that the probability of having zero 
resistant alleles is small in the log-normal model, around $5 \%$, versus
approximately $55 \%$ in our model. 
We believe that the latter is better suited to the considered experiment, 
which consists of a field of small area $F=20 \text{ha}$ 
(the smaller is the area, the larger is the probability of zero resistant allele)
and having no direct nor indirect effect of the target 
herbicide e.g. via seed interchange with other areas subjected to herbicide.

Turning our attention to the comparison between the stochastic and the 
hybrid models (recall $\mu_0$ is identical in both models), 
we consider the period of target herbicide application $t\geq 0$.
The hybrid model  indicates the existence of deterministic bounds to the emergence time and 
visualization index (in fact, $14 \leq t_{\mathfrak{X}} \leq 32$ for all realizations), 
which seems implausible.  
A possible way around this is to ``introduce randomness'' in the parameters 
of the hybrid model, similar to
\cite{Bagavathiannan_etal_2013,Neve_etal_2011,Neve_etal_2003_II,Neve_etal_2011B}.
Even so, we have noted that the allele distribution given by the hybrid 
model evolves in a rather simple fashion along the time 
(see Figure \ref{fig:FreqAlleleAaplicSTOSTO2}), in such a manner that the variance 
remains close to a fixed value for a large time horizon, while it changes dynamically 
in the stochastic model in a fashion that cannot be reproduced by the hybrid model.
\bl{As for the deterministic model, we note that it is more severely affected by 
a lack of randomness, moreover its estimates for the \emph{average} resistant allele frequency 
coincide with the stochastic model in a portion of the time interval only 
(they do not agree in the last three plots of Figure \ref{fig:FreqAlleleAaplicSTOSTO1}).}

\section{Conclusions}
\label{sec:conclusions}

We have proposed a stochastic model for weed control and wild resistance dynamics, which can be used for estimating the wild resistant allele frequency as well as 
its evolution along time, which plays a fundamental role in the weed control. 
The model is individual based, stochastic, and considers some 
phenomena like the relative fitnesses and mutation that are 
prominent in the wild resistance dynamics.


The proposed model brings new insights on the resistance evolution along time 
as presented in Section \ref{sec:result}.
The stochastic model features the largest, dynamic variability of the 
resistant allele frequency estimates and their distributions among all considered models, 
as discussed in more  detail in Section \ref{sec-discussion}.  
The visualization index as defined in \eqref{eq:indexResis} presents 
an abrupt behaviour along time when using the other models, whereas it is 
smooth according our model, which seems more adherent to the case study
considered in the simulations.
This is a relevant index as it may be easily observed in the real-world, 
allowing to test the porposed model in future work.



\appendix

\section{Auxiliary equations and weed model parameters }
\label{sec:app:AE}

\subsection{Allele frequency and seed bank   - $\mu, \eta$}
The frequency of allele $A$ is calculated using
\begin{equation}
\label{eq:AlleleFreq}
\mu(Z_t)  =  \frac{2Z_{1,t} + Z_{2,t}}{\eta(Z_t)},
\end{equation}
for $Z_t\in\mathbb R^3$, 
where $\eta(Z_t) =  Z_{1,t} + Z_{2,t} + Z_{3,t}$. 
Note that $\mu(X_t)$ gives  the frequency
of seed bank, $\mu( Y_t^a)$ the frequency of mature plants
and $\eta(X_t)$ the total amount of seeds at seed bank at time instant $t$.

We define bellow the mean ($\overline{\mu}_{{\bf X}_t}$)  
and standard deviation ($\widehat{\mu}_{{\bf X}_t}$)
for the allele frequency generated by the stochastic model simulation.
We use analogous equations for the seed bank, defined by  
$\overline{\eta}_{{\bf X}_t}$ and $\widehat{\eta}_{{\bf X}_t}$.
\begin{eqnarray*}
\overline{\mu}_{{\bf X}_t}  &=& N^{-1}\sum_{k=1}^{N} \mu({\bf X}^{\omega_k}_{t}),  \\
\widehat{\mu}_{{\bf X}_t}  &=& \overline{\mu}_{{\bf X}_t}  + \sqrt{N^{-1}\sum_{k=1}^{N} \Big(\mu({\bf X}^{\omega_k}_{t}) - \overline{\mu}_{{\bf X}_t} \Big)^2}.  \\
\end{eqnarray*}

\subsection{Resistance visualization - $t_{X}, f_{\bf X}^t$}

We assume that when the number of resistant individuals  is greater than 
$30\%$ of total mature plants, the resistance can be visualized in the field \cite{Christoffoleti_LIVRO_2008}. 
So, we define the  {\it first time of resistance visualization index}
by 
\begin{equation}
\label{eq:indexResis}
t_{X} = \min_{0\leq t \leq T} \{t: J({Y}^a_{t})=1\}, 
\end{equation}
where  $J(Y) = \text{sgn}\left\{\frac{Y_{1} + Y_{2}}{\eta(Y)} - 0.3\right\}$, with  
$sgn$ being the signal function, defined as 
\begin{equation*}
sgn(x)= \Bigg\{ \begin{array}{lr}
        1,   \text{when}  ~ x\geq 0\\
        0,  \text{otherwise.} 
        \end{array}
\end{equation*}
Hence, the accumulated occurrence of resistance visualization can be calculated
as following. For instance, for the stochastic model, we have
\begin{equation}
\label{eq:identResFig}
f_{\bf X}^t = \sum_{k=1}^{N}{\mathbbm{1}}_{t \geq t_{{\bf X}}},
\end{equation}
where $\mathbbm{1}_{\Omega}$ is the indicator function (that is $1$ if $\Omega$
is {\it true} and $0$ if it is {\it false}). Analogous equations are used  
to obtain the accumulated occurrences hybrid ($f_{\mathfrak X}^t$) 
and deterministic ($f_{X}^t$) processes.

\section{Breeding seeds genotype probability - $P(G_t)$}
\label{sec:app:PG}

\par
To obtain the  genotype of the seeds produced by breeding, 
equation \ref{eq:genseed}, 
we need to calculate the probability
that a generated seed has genotype $i$, denoted by $P(G_{i,t})$.
First, we calculate $P(G_{i,t})$  for a generic weed population $W$,
that only make cross-fertilization.
To this end, consider the probability of pick randomly a 
gamete with allele of 
type $A$ produced by this population, denoted by $P(A|W)$. 
Assuming that the mutation may occurs only once per allele during the 
gametogenesis, we have
\[ P(A|W) = P(A^-|\bar m,W )P(\bar m|W) +  P(a^-| m,W)P( m|W),\]
where $A^-$ and $a^-$ are the events of take at random an allele 
at the beginning of gametogenesis,   
and $m$ and $\bar m$ are, respectively, the mutation and non-mutation events. 
Considering the complementarity of the events
$A^-/a^-$ and $\bar m/ m$, and assuming the independence of
mutation event with  allele genotype and $W$, we can write
\begin{equation}
\label{eq:ProbPosGametogenese}
 P(A|W) = P(A^-|W)(1 - 2P(m)) +  P( m).
\end{equation}
%
Then, assuming that the gametes combine randomly to form a seed,
consider to pick at random a seed produced by this subpopulation
for inspection.  Denoting by $G_i$ the event of finding a seed with 
genotype $i$, we have   
\begin{eqnarray}
\label{eqn:PGW1}
P(G_1|W) &=& \left(P(A|W)\right)^2, \\
\label{eqn:PGW2}
P(G_2|W) &=& 2P(A|W)\left(1- P(A|W)\right),\\
\label{eqn:PGW3}
P(G_3|W) &=& \left(1 - P(A|W)\right)^2.
\end{eqnarray}

Now we consider a more general population, in which 
two fertilization mechanism is possible: self and cross-fertilization.
To calculate $P(G_{i,t})$ for this population, we may assume that seeds are
produced by two distinct subpopulations, where each employs only one 
fertilization mechanism.
Hence, using  the total probability rule, we get
\begin{equation}
\label{eq:probTotal}
P(G_{i,t}) = P(G_{i,t}|F_a)P(F_a)+P(G_{i,t}|\bar F_a)P(\bar F_a),
\end{equation}
where $F_a$ is the event of the seed be generated by self-fertilization 
and $\bar F_a$ by cross-fertilization,
and $P(F_a)$ and $P(\bar F_a)$ are,  respectively, its probability of
occurrence.

The self-fertilizing subpopulation may be also divided in 3 subpopulations,
where each has only mature weeds with one genotype of $\mathcal{G}$.
So, using  the total probability rule again, we get
\begin{equation}
\label{eq:probFA}
P(G_{i,t}|F_a) = \sum_{l\in \mathcal{G}} P(G_{i,t}|Y^a_{l},F_a)P(Y^a_{l}),
\end{equation}
where $Y^a_{\ell}$ refers to the event of the weed parent have the 
genotype $\ell$. 
Note that it does not matter to assume either cross or 
self-fertilizing in a sub population with the same genotype. 
What means that we can also use  \eqref{eqn:PGW1}-\eqref{eqn:PGW3} 
for $P(G_{i,t}|Y^a_{l},F_a)$.

Assuming  $P(m)$, $P(F_a)$ and $P(\bar F_a)$ time invariant and given,
to calculate $P(G_{i,t})$, we only need to know $P(Y^a_{\ell})$, and 
$P(G_{i,t}|\bar F_a)$ and $P(G_{i,t}|Y^a_{l},F_a)$ for all $i,l\in \mathcal{G}$.
First, note that the former can be easily obtained doing 
$P(Y^a_{\ell}) = {\bf Y}^{a}_{\ell,t}/{ \eta ({\bf Y}^a_{t})}$ (please, see
\ref{sec:app:AE} for definition of $\eta$).
Now, recalling that $W$ was defined as a general weed 
population,  is sufficient to employ 
$P(A^-|\bar F_a) = \mu({\bf Y}^{a}_{t})$ in \eqref{eq:ProbPosGametogenese}
to obtain $P(G_{i,t}|\bar F_a)$, and  
 $P(A^-|Y^a_{1},F_a) = 1$, $P(A^-|Y^a_{2},F_a) = 0.5$ 
and $P(A^-|Y^a_{3},F_a) = 0$,  to obtain  
 $P(G_{i,t}|Y^a_{l},F_a)$, for all $i,l\in\mathcal G$.

\section{\bf Weed model parameters}
\label{sec-param}
All parameters employed in simulations are presented in Table \ref{ValuePopEco}.
Some parameters (mortality induced by the target herbicide, 
reproduction, self and cross-fertilization probabilities) 
were obtained for the
herbicide nicosulfuron and the life cycle of  \textit{B. pilosa}, 
an aggressive and rather preeminent weed in annual and perennial crops in Brazil. 
According  to greenhouse experiments carried out 
by the Brazilian Agricultural Research Agency  \textit{Embrapa Milho e Sorgo}, 
the recommended dose of nicosulfuron ($60$ $g\,\,ha^{-1}$ for maize cultures)
causes $84.50 \%$ of phytotoxicity in  susceptible \textit{B. pilosa} 
individuals and $22.75 \%$ in resistant individuals. 
According to \cite{SunGardners1990}, we assume a $92\%$ of self-fertilization 
probability.
We set $g = 1846.2$ in order to obtain  $x^{spp}_t = 1500$ when
the density of mature plants  is $1 m^{-2}$, and
$G=70000$, the maximum expected seed production density
(please, see comments below Equation \eqref{eq:meanSeeds}).
We consider a probability of $10^{-9}$ for an allele mutation at 
gametogenesis and an adaptive cost of $3\%$
following \cite{Neve_2008} and \cite{Roux_etal2008}, respectively.
For the remaining parameters we adopt the suggestions of specialists.
\par

\begin{table}[!h]
\begin{center}
\caption{Parameters used in the numeric simulations}
\label{ValuePopEco}
\begin{tabular}{l c l l c} \hline
Parameter          &  Value  & & Parameter          &  Value \\ \hline 
$F $                     & $2\times 10^5$ & &$P(m) $                     & $ 10^{-9}$ \\ 
$\delta $                & $0.10$  & &$P(F_a)$               & $0.92$  \\
$\gamma$                 & $0.10$  & &$g$                      & $1846.2$\\ 
$\psi $                  & $0.14$  & &$G$                      & $70000$ \\
$c$                      & $0.03$  & &$\rho^{S}(u^{*})$           &  $0.8450$  \\ 
$\kappa $                & $0.10$  & &$\rho^{R}(u^{*})$           &   $0.2275$  \\
$\varphi$                & $0.20$  & &$\rho(v^*)$           &  $0.8450$  \\
\hline 
\end{tabular}
\end{center}
\end{table}


\newpage
\section*{Acknowledgement}
This work was supported by EMBRAPA, CAPES, FAPESP under Grant 13/19380-8, 
CNPq under Grants 306466/2010 and 311290/2013-2 and CNPq Grant 158297/2013-0.

\section*{References}
\bibliographystyle{elsarticle-num} 
\bibliography{WeedManagement}

\begin{thebibliography}{10}
\expandafter\ifx\csname url\endcsname\relax
  \def\url#1{\texttt{#1}}\fi
\expandafter\ifx\csname urlprefix\endcsname\relax\def\urlprefix{URL }\fi
\expandafter\ifx\csname href\endcsname\relax
  \def\href#1#2{#2} \def\path#1{#1}\fi

\bibitem{Jasieniuk_1996}
M.~Jasieniuk, A.~BruleBabel, I.~Morrison, {The evolution and genetics of
  herbicide resistance in weeds}, {Weed Science} {44}~({1}) ({1996})
  {176--193}.

\bibitem{Renton_etal2014}
M.~Renton, R.~Busi, P.~Neve, D.~Thornby, M.~Vila-Aiub, Herbicide resistance
  modelling: past, present and future, Pest Management Science 70~(9) (2014)
  1394--1404.

\bibitem{Holst_etal_2007}
N.~Holst, I.~A. Rasmussen, L.~Bastiaans, Field weed population dynamics: a
  review of model approaches and applications, Weed Research 47~(1) (2007)
  1--14.

\bibitem{Bagavathiannan_etal_2013}
M.~V. Bagavathiannan, J.~K. Norsworthy, K.~L. Smith, P.~Neve, {Modeling the
  Evolution of Glyphosate Resistance in Barnyardgrass (Echinochloa crus-galli)
  in Cotton-Based Production Systems of the Midsouthern United States}, {Weed
  Technology} {27}~({3}) ({2013}) {475--487}.

\bibitem{Neve_etal_2011}
P.~Neve, J.~K. Norsworthy, K.~L. Smith, I.~A. Zelaya, Modelling evolution and
  management of glyphosate resistance in amaranthus palmeri, Weed Research
  51~(2) (2011) 99--112.

\bibitem{Thornby_etal_2009}
D.~F. Thornby, S.~R. Walker, {Simulating the evolution of glyphosate resistance
  in grains farming in northern Australia}, {Annals of Botany} {104}~({4})
  ({2009}) {747--756}.

\bibitem{Cavan_2000}
G.~Cavan, J.~Cussans, S.~R. Moss, Modelling different cultivation and herbicide
  strategies for their effect on herbicide resistance in alopecurus
  myosuroides, Weed Research 40~(6) (2000) 561--568.

\bibitem{Diggle_etal_2003}
A.~J. Diggle, P.~B. Neve, F.~P. Smith, Herbicides used in combination can
  reduce the probability of herbicide resistance in finite weed populations,
  Weed Research 43~(5) (2003) 371--382.

\bibitem{Neve_etal_2003_I}
P.~Neve, A.~Diggle, F.~Smith, S.~Powles, {Simulating evolution of glyphosate
  resistance in Lolium rigidum I: population biology of a rare resistance
  trait}, {Weed Research} {43}~({6}) ({2003}) {404--417}.

\bibitem{Neve2008}
P.~Neve, Simulation modelling to understand the evolution and management of
  glyphosate resistance in weeds, Pest Management Science 64~(4) (2008)
  392--401.

\bibitem{Renton2011}
M.~Renton, A.~Diggle, S.~Manalil, S.~Powles, Does cutting herbicide rates
  threaten the sustainability of weed management in cropping systems?, Journal
  of Theoretical Biology 283~(1) (2011) 14 -- 27.

\bibitem{Manalil2012}
S.~Manalil, M.~Renton, A.~Diggle, R.~Busi, S.~B. Powles, Simulation modelling
  identifies polygenic basis of herbicide resistance in a weed population and
  predicts rapid evolution of herbicide resistance at low herbicide rates, Crop
  Protection 40~(0) (2012) 114 -- 120.

\bibitem{Neve_etal_2003_II}
P.~Neve, A.~Diggle, F.~Smith, S.~Powles, {Simulating evolution of glyphosate
  resistance in Lolium rigidum II: past, present and future glyphosate use in
  Australian cropping}, {Weed Research} {43}~({6}) ({2003}) {418--427}.

\bibitem{Neve_etal_2011B}
P.~Neve, J.~K. Norsworthy, K.~L. Smith, I.~A. Zelaya, {Modeling Glyphosate
  Resistance Management Strategies for Palmer Amaranth (Amaranthus palmeri) in
  Cotton}, {Weed Technology} {25}~({3}) ({2011}) {335--343}.

\bibitem{Smith_Book_1998}
J.~M. Smith, Evolutionary Genetics, Oxford University Press, 1998.

\bibitem{Radosevich_etal_BOOK_2007}
C.~M.~G. Steven R.~Radosevich, Jodie S.~Holt, Ecology of Weeds and Invasive
  Plants: Relationship to Agriculture and Natural Resource Management., John
  Wiley \& Sons, 2007.

\bibitem{Zimdahl_BOOK_2013}
R.~L. Zimdahl, Fundamentals of Weed Science, Fourth Edition, Elsevier, 2013.

\bibitem{degroot2002probability}
M.~DeGroot, M.~Schervish, Probability and Statistics, Addison-Wesley series in
  statistics, Addison-Wesley, 2002.

\bibitem{RouxReboud2007}
F.~Roux, X.~Reboud, Herbicide resistance dynamics in a spatially heterogeneous
  environment, Crop Protection 26~(3) (2007) 335 -- 341, weed Science in Time
  of Transition.

\bibitem{Roux_etal2008}
F.~Roux, M.~Paris, X.~Reboud, {Delaying weed adaptation to herbicide by
  environmental heterogeneity: a simulation approach}, {Pest Management
  Science} {64}~({1}) ({2008}) {16--29}.

\bibitem{Neve_2008}
P.~Neve, {Simulation modelling to understand the evolution and management of
  glyphosate resistant in weeds}, {Pest Management Science} {64}~({4}) ({2008})
  {392--401}.

\bibitem{Christoffoleti_LIVRO_2008}
P.~J.~C. (coordenador), Aspectos de Resistência de Plantas Daninhas a
  Herbicidas, 3rd Edition, Associação Brasileira de Ação à Resistência de
  Plantas aos Herbicidas (HRAC-BR), 2008.

\bibitem{SunGardners1990}
M.~Sun, F.~Ganders, {Outcrossing Rates and Allozyme Variation in Rayed And
  Rayless Morphs of Bidens-pilosa}, {Heredity} {64}~({1}) ({1990}) {139--143}.

\end{thebibliography}

\end{document}